\documentclass[sigconf,authorversion,nonacm]{acmart}

\AtBeginDocument{%
  \providecommand\BibTeX{{%
    \normalfont B\kern-0.5em{\scshape i\kern-0.25em b}\kern-0.8em\TeX}}}

\setcopyright{acmcopyright}
\copyrightyear{2020}
\acmYear{2020}
\acmDOI{10.1145/3377811.3380405}

\acmConference[ICSE '22]{44th International Conference on Software Engineering}{May 21--29, 2022}{Pittsburgh, USA}
\acmBooktitle{44th International Conference on Software Engineering (ICSE '22), May 21--29, 2022, Pittsburgh, USA}
\acmPrice{15.00}
\acmISBN{XXXX}

\usepackage{amsmath,amsfonts}
\usepackage{algorithm}
\usepackage[noend]{algpseudocode}
\usepackage{textcomp}
\usepackage{soul}
\usepackage{booktabs}
\usepackage[utf8]{inputenc}
\usepackage[T1]{fontenc}
\usepackage{microtype}
\usepackage{rotating}
\usepackage{graphicx}
\usepackage{paralist}
\usepackage{tabularx}
\usepackage{multicol}
\usepackage{multirow}
\usepackage{pbox}
\usepackage{enumitem}	
\usepackage{colortbl}
\usepackage{pifont}
\usepackage{xspace}
\usepackage{url}
\usepackage{tikz}
\usepackage{tabularx}
\usepackage{fontawesome}
\usepackage{lscape}
\usepackage{listings}
\usepackage{color}
\usepackage{anyfontsize}
\usepackage{comment}
\usepackage{soul}
\usepackage{multibib}
\usepackage{tcolorbox}

\newcommand{\ie}{\emph{i.e.,}\xspace}
\newcommand{\eg}{\emph{e.g.,}\xspace}
\newcommand{\etc}{etc.\xspace}
\newcommand{\etal}{\emph{et~al.}\xspace}
\newcommand{\secref}[1]{Section~\ref{#1}\xspace}
\newcommand{\figref}[1]{Fig.~\ref{#1}\xspace}

\newcommand{\tabref}[1]{Table~\ref{#1}\xspace}

\newtcolorbox{resultbox}{boxrule=1pt,leftrule=5pt,arc=0pt,auto outer arc, top=0.5mm, 
bottom=0.5mm, left=1mm, right=1mm}
\title{Using Pre-Trained Models to Boost Code Review Automation}

\author{Rosalia Tufano}
\affiliation{%
  \institution{\small{SEART @ Software Institute\\Universit\`a della Svizzera italiana}}
 \country{Switzerland}
}
\author{Simone Masiero}
\affiliation{%
  \institution{\small{SEART @ Software Institute\\Universit\`a della Svizzera italiana}}
  \country{Switzerland}
}
\author{Antonio Mastropaolo}
\affiliation{%
  \institution{\small{SEART @ Software Institute\\Universit\`a della Svizzera italiana}}
 \country{Switzerland}
}
\author{Luca Pascarella}
\affiliation{%
  \institution{\small{SEART @ Software Institute\\Universit\`a della Svizzera italiana}}
  \country{Switzerland}
}
\author{Denys Poshyvanyk}
\affiliation{%
  \institution{\small{SEMERU @ Computer Science Department\\William and Mary}}
 \country{USA}
}
\author{Gabriele Bavota}
\affiliation{%
  \institution{\small{SEART @ Software Institute\\Universit\`a della Svizzera italiana}}
  \country{Switzerland}
}


\begin{document}

\begin{abstract}
Code review is a practice widely adopted in open source and industrial projects. Given the non-negligible cost of such a process, researchers started investigating the possibility of automating specific code review tasks. We recently proposed Deep Learning (DL) models targeting the automation of two tasks: the first model takes as input a code submitted for review and implements in it changes likely to be recommended by a reviewer; the second takes as input the submitted code and a reviewer comment posted in natural language and automatically implements the change required by the reviewer. While the preliminary results we achieved are encouraging, both models had been tested in rather simple code review scenarios, substantially simplifying the targeted problem. This was also due to the choices we made when designing both the technique and the experiments. In this paper, we build on top of that work by demonstrating that a pre-trained Text-To-Text Transfer Transformer (T5) model can outperform previous DL models for automating code review tasks. Also, we conducted our experiments on a larger and more realistic (and challenging) dataset of code review activities.
\end{abstract}

\maketitle

\section{Introduction} \label{sec:intro}

The benefits of code reviews have been widely recognized, with several studies providing evidence of the higher quality of reviewed code \cite{McIntosh:msr2014,morales2015saner,Bavota:icsme2015}. Also, code reviews help in preventing bugs and foster knowledge transfer among developers  \cite{Caitlin:icse2018,bacchelli2013expectations}. However, studies on code reviews also highlighted an additional cost that such a process entails: Empirical evidence suggests that large software projects can undergo hundreds of code reviews per month. This applies to both open-source (\eg $\sim$500 reviews per month in Linux \cite{Rigby:tosem2014}) and industrial (\eg $\sim$3k reviews per month in Microsoft Bing \cite{Rigby:fse2013}) projects. As a result, developers can spend many hours per week reviewing code \cite{Bosu:2013}. \eject

Given the non-negligible cost of code review, we recently proposed the automation of specific code review tasks: The goal is not to replace developers, but to help them save time in two scenarios. The first is that of a contributor (\ie the developer submitting the code for review) who wants to receive a rapid feedback about the code they wrote before submitting it for review. The feedback is provided by a Deep Learning (DL) model trained to take as input the code to submit for review $C_s$ and provide as output a revised version of $C_s$ (\ie $C_r$) implementing code changes that are likely to be recommended by a reviewer. 

The second scenario concerns the reviewer(s) involved in the process: a DL model is trained to take as input (i) the code $C_s$ submitted for review, and (ii) a comment $R_{nl}$ written by the reviewer in natural language to request a specific change on $C_s$. The output of the model is a revised version of $C_s$ (\ie $C_r$) implementing the changes recommended in $R_{nl}$. The idea here is that the reviewer can use the model to provide the contributor with a concrete example of the code changes that they would like to see implemented.

In our previous work \cite{Tufano:icse2021} we trained and experimented with the DL models on a dataset composed of $\sim$17k triplets $\langle$$C_s$, $R_{nl}$, $C_r$$\rangle$ extracted from code reviews performed in GitHub \cite{GitHub} and Gerrit \cite{Gerrit}. In particular, the model recommending code changes to the contributor is an encoder-decoder model with one encoder taking $C_s$ as input and one decoder generating $C_r$. Our evaluation shows that this model can recommend a change as a reviewer would do in 3\% (single prediction) to 16\% of the cases (10 different predictions). The model employed in the second scenario (\ie the automated implementation of a comment recommended by the reviewer), has instead two encoders taking as input $C_s$ and $R_{nl}$, respectively, and one decoder generating $C_r$. This model can successfully implement a change recommended by a reviewer in 12\% (single prediction) to 31\% (10 different predictions) of the cases.

While these results represent our first step towards automating code review tasks, our approach \cite{Tufano:icse2021} as well as the conducted empirical study suffers of several limitations we try to overcome in this paper. First, we adopted a code abstraction process to reduce the vocabulary size and simplify the learning of the DL model. This means that the model did not work on the raw source code, but on an abstracted version of it in which, for example, variable identifiers were replaced with a special \texttt{VAR\_ID} token, where \texttt{ID} is a progressive number (\eg the second declared variable is represented by \texttt{VAR\_2}). The possibility to go back to raw source code was guaranteed by keeping a map linking abstracted to raw tokens in $C_s$ (\eg \texttt{VAR\_1} $\rightarrow$ \texttt{i}). 

While such a procedure simplifies the learning of the model, it poses a strong limitation on the variety of code review tasks that can be supported by such a model. Indeed, the abstraction process forces to exclude from the dataset of triplets $\langle$$C_s$, $R_{nl}$, $C_r$$\rangle$ all those in which $C_r$ introduces identifiers or literals that were not present in $C_s$. This is necessary because the abstraction map is built on $C_s$ and, if a new variable \texttt{VAR\_2} is introduced in $C_r$ during the review process, such a variable cannot be mapped back to raw source code, making such an approach unusable in practice. This means that the triplets $\langle$$C_s$, $R_{nl}$, $C_r$$\rangle$ on which we evaluated our approach \cite{Tufano:icse2021} were relatively simple changes implemented during code review, not requiring the introduction of new identifiers or literals.

Second, to simplify the learning, we only considered triplets $\langle$$C_s$, $R_{nl}$, $C_r$$\rangle$ in which both the code submitted for review ($C_s$) and the revised code ($C_r$) had no more than 100 tokens \cite{Tufano:icse2021}. Again, this reduced the complexity of the tackled problem.

Basically, the two above choices resulted in training and experimenting the proposed models on quite simple code review instances only representative of a minority of the code transformations actually implemented during code reviews. 

In this paper, we build on top of our previous work \cite{Tufano:icse2021} experimenting with DL models for code review automation in more realistic and challenging scenarios. We start by training the recently proposed Text-To-Text-Transfer Transformer (T5) model \cite{raffel2019exploring} on a dataset similar to the one used in \cite{Tufano:icse2021}. However, we adopt a tokenizer (\ie SentencePiece \cite{Kudo:sentencePiece}) that allows us to work with raw source code, without the need for code abstraction. Also, we increase the maximum length of the considered code components from 100 ``abstracted'' tokens to 512 ``SentencePiece'' tokens (\ie $\thicksim$390 ``abstracted'' tokens). The absence of an abstraction mechanism and the increased upper bound for input/output length allowed us to build a substantially larger dataset as compared to the one used in \cite{Tufano:icse2021} (168k instances \emph{vs.} 17k) and, more importantly, to feature in such a dataset a wider variety of code transformations implemented in the code review process, including quite challenging instances such as those requiring the introduction of new identifiers and literals (accounting for 63\% of the new dataset we built). Also, we experimented with the automation of a third task related to the code review process: Given the code submitted for review ($C_s$), generating a natural language comment $R_{nl}$ requesting to the contributor code changes as a reviewer would do (\ie simulating a reviewer commenting on the submitted code).

We also compare the T5 model with the encoder-decoder model presented in our previous work on the original dataset used in \cite{Tufano:icse2021}. Our results show the superior performance of T5, which represents a significant step forward in automating code review tasks.


\noindent To summarize, the \textbf{contributions} of this work are:

(i) A novel approach for code review automation overcoming several limitations of the state-of-the-art technique \cite{Tufano:icse2021};

(ii) A comprehensive empirical evaluation of such an approach, including a comparison with our previous technique \cite{Tufano:icse2021};

(iii) The automation of a third task: Given the code submitted for review, automatically generating natural language comments requesting changes as reviewers would do;

(iv) A code review dataset to train and test DL models in more realistic scenarios as compared to the one used in \cite{Tufano:icse2021};

(v) A comprehensive replication package \cite{replication}.
\section{T5 to Automate Code Review} \label{sec:t5}
We describe the DL model we adopt, the construction process of the datasets needed for its training, and the procedure used for hyperparameter search, model training, and generation of predictions.

\subsection{Text-to-Text Transfer Transformer (T5)}
The Text-to-Text Transfer Transformer, or simply T5, is not merely a model. Raffel \etal \cite{raffel2019exploring} compare ``\emph{pre-training objectives, architectures, unlabeled data sets, transfer approaches, and other factors on dozens of language understanding tasks}''. 

The result of this exploration is the best combination of architectures and training techniques, namely T5. T5 is based on the Transformer \cite{vaswani2017attention} architecture. The proposed implementation differs only in some details (regarding the normalization layer and the embedding scheme) from its original form. Raffel \etal proposed several versions of T5, differing from each other in their size (\eg number of layers) and, as a consequence, training complexity. In this work we adopt the \emph{small} version of T5 consisting of: 8-headed attention, 6 layers in both the encoder and the decoder, each having a dimensionality of 512 and the output dimensionality of 2,048 ($\sim$ 60M parameters).

The model is subjected to a first training (pre-training) whose purpose is to provide it with a general knowledge useful to solve a set of related tasks. Suppose, for example, that we want to train a model able to (i) translate English to German, and (ii) summarize English text. Instead of starting by training the model for these two tasks, T5 can be pre-trained in an unsupervised manner by using the \emph{denoising objective} (or \emph{masked language modeling}): The model is fed with sentences having 15\% of their tokens (\eg words in English sentences or code tokens in Java statements) randomly masked and it is asked to predict them. By learning how to predict the masked tokens, the model can acquire general knowledge about the language of interest. In our example, we could pre-train the model on English and German sentences.

Once pre-trained, T5 is fine-tuned on the downstream tasks in a supervised fashion. Each task is formulated in a ``text-to-text'' format (\ie both the input and the output of the model are represented as text). For example, for the translation task a dataset composed of pairs of English and German sentences allows to fine-tune the model. Similarly, the summarization task requires the input English text and a corresponding summary. In the next sections we explain how we pre-train and fine-tune T5 to support code review tasks. 

\subsection{Training Data}
We describe the process used to build the datasets needed for the pre-training (\secref{sub:pretraining}) and fine-tuning (\secref{sub:finetuning}) of T5. Part of the fine-tuning dataset has been used for hyperparameter search (\secref{sub:hyper}) and for testing the performance of T5 (\secref{sec:design}).

\subsubsection{Pre-training Dataset} \label{sub:pretraining}

Given the goal of the pre-training phase (\ie providing the model with general knowledge about the languages of the downstream tasks) we built a dataset allowing to train T5 on Java and technical English. 

Indeed, besides source code, technical English is instrumental in a code review process in which reviewers post natural language comments about code. 

We start from two datasets featuring instances including both source code and technical English: the official Stack Overflow dump (SOD) \cite{so_dump} and CodeSearchNet (CSN) \cite{Java:CodeSearchNet}. Stack Overflow is a Q\&A website for programmers. The data dump we used collects all the questions and relative answers between 2006 and 2020 for a total of roughly 51M posts (where a post is a single question or answer). A post includes English text (as per the SO guidelines) and/or code snippets. Posts are usually accompanied by tags characterizing their topic (\eg \emph{Java}, \emph{Android}) and can be rated with \emph{up-}/\emph{down-votes} and, for what concerns the answers, they can be marked as the ``accepted answer'' from the question's author. 

We extracted from the SOD all the answers (i) having a \emph{Java} tag; (ii) containing at least one \texttt{<pre><code>} HTML tag to ensure the presence of at least one code snippet in the answer; and (iii) having at least 5 up-votes and/or being the accepted answer. These filters are justified by the goal of our pre-training. Indeed, we want the model to acquire knowledge about technical English and Java: focusing on answers containing at least one code snippet increases the chances that their natural language text refers to an implementation task, similarly to what happens in code review. Also, the up-votes/accepted answer filter aims at discarding low-quality instances containing, for example, wrong code solutions. This is also the reason why we focused on high-quality answers likely to contain working solutions rather than on questions that, even if up-voted (\eg because they are relevant for many users) may contain wrong implementations. From this step we obtained 1,018,163 candidate instances from the SOD. 

On each selected answer $a$, we performed the following cleaning steps: We remove emojis, non-latin characters, control characters, trailing spaces and multiple white spaces. Some special symbols are replaced using latin characters having the same meaning, \eg "$\ge$" is replaced with "$> =$". Moreover, we replace any embedded link with a special tag "\texttt{<LINK\_i>}", with \texttt{$i$} being an integer ranging from 0 to $n-1$, where $n$ is the number of links in $a$. Finally, we removed all the instances having less than ten tokens or more than 512 (40,491). This left us with 977,379 valid instances. 

The CSN \cite{Java:CodeSearchNet} Java dataset features ~1.5M unique Java methods, some of which containing their Javadoc. We filtered out all those in which a Javadoc was not available or it did not contain any letter, removing 1,034,755 of them. Unlike the SOD, CSN can contain instances in which the ``textual part'' (\ie the method comment) is not in English. To partially address this issue, we exclude pairs in which no Latin characters were found. While this does not exclude all non-English comments, at least identifies and removes those written in specific languages (\eg Russian, Chinese) (15,229). We decided to accept some level of noise in the pre-training dataset (\eg comments written in French) since (i) given the size of this dataset, this little amount of noise should not substantially affect the model's performance, and (ii) the pre-training dataset is not used as test set to assess the performance of the approach. As we will explain later, a more fine-grained cleaning has been performed for the fine-tuning dataset that, instead, is used for performance evaluation.
On the 519,905 remaining instances, we performed the same cleaning steps described for the SOD (\eg remove emojis). Finally, from each pair we obtain a single string concatenating the Javadoc comment and the code, retaining the ones having more than ten and less than 512 tokens (507,947 instances left).

By putting together the instances collected from the SOD and CSN we obtained the pre-training dataset consisting of 1,485,326 instances. To perform the pre-training, we randomly mask in each instance 15\% of its tokens. The masked tokens are replaced with \emph{sentinel tokens} \texttt{<extra\_id\_i>}, where \texttt{$i$} is an increasing number ranging from 0 up to $n-1$, where $n$ is the number of tokens masked in a given instance. If several contiguous tokens are masked they are replaced by a single sentinel token. These ``masked instances'' represent the input of the model during the pre-training. The target (\ie the string the model is expected to generate) is built concatenating the sentinel tokens and the token(s) they are masking. An extra sentinel token is added to indicate the end of the string. 

Our pre-training dataset is publicly available \cite{replication}. 

\subsubsection{Fine-tuning Datasets} \label{sub:finetuning}

To create the fine-tuning dataset we mined Java open source projects from GitHub using the web application by Dabic \etal~\cite{dabic2021sampling}. Using the querying interface~\cite{ghs}, we selected all Java projects having at least 50 pull requests (PRs), ten contributors, ten stars, and not being forks. The filters aim at (i) ensuring that enough ``code review'' material is contained in the projects (\ie at least 50 PRs); (ii) discarding personal/toy projects (at least ten contributors and stars); and (iii) reducing the chance of mining duplicated code. This resulted in a list of 4,901 projects. We also mined the six Gerrit \cite{Gerrit} installations used in \cite{Tufano:icse2021} containing code review data about 6,388 projects. 

From both the GitHub and the Gerrit datasets we extract triplets $<m_s, c_{nl}, m_r>$, where $m_s$ is a method submitted for the review; $c_{nl}$ is a single reviewer's comment suggesting code changes for $m_s$; and $m_r$ is the revised version of $m_s$ implementing the reviewer's recommendation $c_{nl}$. Note that (i) we only looked for PRs that are accepted at the end of the code review, since we want to learn how to recommend changes that, at the end, can lead to code considered good from a reviewer's perspective; and (ii) a single PR in GitHub and Gerrit can result in several triplets for our dataset. Indeed, we mine the different review rounds in each PR. For example, a method $m_s$ can be submitted for review, receiving a comment $c_{nl}$ asking for changes (first round). The revised version of $m_s$ addressing $c_{nl}$ is then resubmitted ($m_r$), resulting in the second review round (possibly leading to additional comments and revisions of the method). We stop when the code is formally accepted. 

Overall, we mined 382,955 valid triplets from GitHub and Gerrit using the pipeline from \cite{Tufano:icse2021} that we summarize in the following (see \cite{Tufano:icse2021} for additional details). We target triplets in which a comment $c_{nl}$ has been posted by a \textbf{reviewer} on a method $m_s$. We can identify these cases since both GitHub and Gerrit (i) provide information about the developers submitting the code and posting comments in the review process; and (ii) allow to retrieve the specific code line(s) $c_{nl}$ refers to (\ie the code in $m_s$ that has been highlighted by the reviewer when posting the comment). 

We exclude all the comments posted by the authors of the code (\eg to reply to reviewers), since they do not represent a review of the code. Thus, the triplets in our dataset have $c_{nl}$ being a single comment posted by a reviewer. Also, we exclude $c_{nl}$ linked to inline comments (rather than code lines) in $m_s$, since we target the fixing of code-related issues. To consider a triplet as valid, $c_{nl}$ must be the only comment posted by a reviewer on $m_s$ in that specific review round. 

In this way, we can be confident that the revised version submitted later on by the author ($m_r$) actually aimed at implementing $c_{nl}$. Also, $m_r$ must differ from $m_s$ (\ie a change must have been implemented in the code to address $c_{nl}$). From the technical point of view, the parsing of the methods from the patches submitted for review has been done using the lizard library \cite{Lizard}. Note that, the removal of triplets in which $c_{nl}$ include more than one comment has been done later in the processing pipeline (we will get back to this point). Indeed, before we had to clean comments possibly just representing noise.

As done for the pre-training dataset, we performed some cleaning steps. We replaced any link with the numbered token \texttt{<LINK\_i>}, with \texttt{$i$} being an integer ranging from 0 to $n-1$, where $n$ is the total number of links in $c_{nl}$, $m_s$ and $m_r$. If the same link appears in different parts (\eg in $c_{nl}$ and $m_r$), it is replaced with the same token. We also removed any emoji and non-ascii characters from the comments, extra spaces and control characters from both the comments and the methods, and inline comments from the methods (we are not interested in addressing issues related to internal comments).

After the cleaning process we obtained some triplets in which $c_{nl}$ became an empty string or where $m_s$ and $m_r$ became equal (\eg they only differed for some spaces before the cleaning). We removed these instances (-33,005) as well as those  having $c_{nl}$ + $m_s$ or $m_r$ longer than 512 tokens (-61,233). We considered the sum of $c_{nl}$ and $m_s$ in terms of length because, for one of the tasks (\ie the automated implementation of a comment posted by a reviewer), they will be concatenated to form the input for the model.

Then, we removed from our triplets non-relevant comments (-28,581), \ie comments not recommending code change suggestions (\eg ``looks good to me''). In \cite{Tufano:icse2021} we manually crafted a set of natural language patterns to spot non-relevant comments (\eg single-word comments containing words such as ``thanks'', ``nice'', \etc). We have extended this set since we noticed that in our richer dataset several non-relevant comments were left by these patterns. Such analysis has been done by one of the authors by manually inspecting all the triplets having $c_{nl}$ consisting of less than six words. The updated heuristics are available in our replication package \cite{replication}.

We also excluded triplets including non-English $c_{nl}$ comments (-4,815) through a pipeline composed by three language detector tools. A preliminary classification has been performed using the Python libraries langdetect \cite{langdetect} and pycld3 \cite{pycld3}. If both of these tools classify the comment as non-English, we relied on the Google language detection API for a final decision. Such a process was needed since we noticed that the Google API was the most accurate in detecting the language, especially when the comments also featured code constructs in them. In this scenario, the Python libraries often generated false negatives (\ie classifying an English sentence as non-English). However, we had a limited number of requests available for the Google API. Thus, we performed a pre-filtering using the Python libraries and, when they both reported the comment as being not in English, we double checked using the Google API. 

After this cleaning process, we excluded all triplets featuring more than one comment in $c_{nl}$ (-86,604). Finally, we removed all the duplicates from the fine-tuning dataset (-918). To be conservative, we identify as duplicates two triplets having the same $m_s$ (thus, even triplets having the same $m_s$ but different $c_{nl}$/$m_r$ have been removed). 

The resulting dataset features 167,799 triplets that have been used to build the three fine-tuning datasets needed for the three tasks we aim at automating. In the first task (\emph{code-to-code}) the model takes as input $m_s$ with the goal of automatically generating its revised version $m_r$, implementing code changes that may be required in the code review process. Thus, the fine-tuning dataset is represented by pairs $m_s \rightarrow m_r$.

In the second task (\emph{code\&comment-to-code}) the model takes as input both $m_s$ and a comment $c_{nl}$ posted by the reviewer and targets the generation of $m_r$, the revised version of $m_s$ implementing the code changes recommended in $c_{nl}$. 

The $m_s$  code contains two special tags \texttt{<START>}, \texttt{<END>} marking the portion of the code $c_{nl}$ refers to. The fine-tuning dataset of this second task is represented by pairs $<m_s, c_{nl}> \rightarrow m_r$.

Finally, in the third task (\emph{code-to-comment}) the model takes as input $m_s$ and aims at generating a natural language comment ($c_{nl}$) suggesting code changes as a reviewer would do. The fine-tuning dataset is represented by pairs $m_s \rightarrow c_{nl}$.

\begin{table}[h!]
    \scriptsize
        \caption{Pre-training and fine-tuning datasets (\# instances)\vspace{-0.2cm}}
        \begin{tabular}{lrrr}
                \toprule
                \textbf{Dataset} & \textbf{train} & \textbf{evaluation} & \textbf{test}\\\midrule
                \textbf{Pre-training} &   &  & \\
                \hspace{3mm}\emph{Stack Overflow} & 977,379  & - & -\\
                 \hspace{3mm}\emph{CodeSearchNet}& 507,947  & - & -\\
                \textbf{Fine-tuning} &   134,239  & 16,780 & 16,780\\
                \bottomrule
        \end{tabular}
    \label{tab:datasets}
\end{table}

All three fine-tuning datasets have been split into 80\% training, 10\% evaluation, and 10\% test. \tabref{tab:datasets} summarizes the number of instances in the datasets: The pre-training is only used for training, while the fine-tuning datasets are exploited also for the hyperparameter tuning (\emph{evaluation}) and for assessing the performance of the model (\emph{test}). In \tabref{tab:datasets} we only report information for a single fine-tuning dataset (rather than for the three previously described), since all three fine-tuning datasets contain the same number of instances. Indeed, they are all derived from the same set of triplets.

\subsection{Training and Hyperparameter Search} \label{sub:hyper}

Raffel \etal \cite{raffel2019exploring} showed the major role pre-training plays on the performance of T5 models. The importance of pre-training has also been confirmed (for other Transformer-based models) in the context of code-related tasks such as test case generation \cite{tufano2020unitTest}. To further study this aspect, we decided to experiment with both a pre-trained and a non pre-trained model, both of which have been subject to a hyperparameter tuning process.

Since we adopted the \emph{small} version of T5 presented by Raffel \etal \cite{raffel2019exploring}, we did not experiment with variations related to its architecture (\eg changing the number of layers or the number of hidden units). Though, as also done by Mastropaolo \etal \cite{mastropaolo2021studying}, we experimented with different learning rate configurations: (i) \emph{Costant Learning Rate} (C-LR), in which the learning rate value is fixed during the training; (ii) \emph{Inverse Square Root Learning Rate} (ISR-LR), in which the learning rate value decays as the inverse square root of the training step; (iii) \emph{Slanted Triangular Learning Rate} (ST-LR) in which first the learning rate linearly increases and then it linearly decays returning to the starting value; (iv) \emph{Polynomial Decay Learning Rate} (PD-LR), in which the learning rate polynomially decays to a fixed value in a given number of steps. 

The hyperparameter tuning has been done for the fine-tuning phase only. Indeed, even though we just focus on one hyperparameter, such a process still remains quite expensive, requiring the training of eight different T5 models (\ie pre-trained and non pre-trained each with four different learning rates). 

For pre-training we use the same configuration proposed by Raffel \etal in \cite{raffel2019exploring}. We pre-trainied the model on the pre-training dataset (\tabref{tab:datasets}) for 200k steps ($\sim$34 epochs). Starting from the pre-trained model, we fine-tuned for 75k steps four different models, each using one of the experimented learning rates. 

Since the goal of this procedure is to find the best learning rate for the three code review tasks, we fine-tuned each of these models using a mixture of the three tasks: A single model is trained to support all three tasks using the union of their training sets. This is one of the characteristics of T5, the possibility to train a single model for multiple tasks. The same approach has been used for the non pre-trained model: In this case four T5 models (one per learning rate) have been directly fine-tuned. 

We assessed the performance of the eight models on the evaluation set of each task in terms of ``perfect predictions'', namely cases in which the generated output was identical to the target (expected) string. \tabref{tab:hyper} reports the achieved results. As it can be seen, no learning rate achieves the best results in all the tasks. Nevertheless, ST-LR shows better overall performance and, for this reason, is the one we adopt in our experiments.

\begin{table}[h!]
    \scriptsize
        \caption{Hyperparameter tuning results}
        \begin{tabular}{lrrrr}
                \toprule
                \multicolumn{1}{c}{\textbf{Task}} & \multicolumn{4}{c}{\textbf{Learining Rate Strategy}} \\\cline{2-5}
                & \textbf{C-LR} & \textbf{ISR-LR} & \textbf{ST-LR} & \textbf{PD-LR} \\\midrule
                \multicolumn{5}{c}{\textbf{Pre-Trained}}\\\cline{1-5}
                \textbf{code-to-code} & 2.68\% & 3.68\% & \textbf{4.64\%} & 2.53\% \\
                \textbf{code\&comment-to-code} & \textbf{10.39\%} & 9.23\% & 8.46\% & 9.89\% \\
                \textbf{code-to-comment} & 0.15\% & 0.32\% & \textbf{0.60\%} & 0.15\% \\\midrule
                \multicolumn{5}{c}{\textbf{Non Pre-Trained}}\\\cline{1-5}
                \textbf{code-to-code} & 1.23\% & 3.71\% & \textbf{4.16\%} & 1.22\% \\
                \textbf{code\&comment-to-code}  & 5.05\% & \textbf{6.41\%} & 6.24\% & 5.18\% \\
                \textbf{code-to-comment} & 0.09\% & 0.44\% & \textbf{0.49\%} & 0.03\% \\
                \bottomrule
        \end{tabular}
    \label{tab:hyper}
\end{table}

Given the best configuration for both the pre-trained and the non pre-trained models, we fine-tuned them for a maximum of 300k steps using an \emph{early stop strategy}. This means that we saved a checkpoint of the model every 10k steps computing its performance in terms of ``perfect predictions'' on the evaluation set and stopped the training if the performance of the model did not increase for three consecutive checkpoints (to avoid overfitting). 

\subsection{Generating Predictions} \label{sub:generation}
Once the models are trained, they can be used to generate predictions. As done in previous work, we adopt a beam search strategy \cite{Raychev:pldi2014} to generate multiple predictions given a single input. For example, in the case of the \emph{code-to-code} task, for a single $m_s$ method provided as input multiple $m_r$ candidates can be generated. When we ask the model to generate $k$ predictions, it generates the $k$ most probable sequences of tokens given the input sequence; $k$ is known as the \emph{beam size} and we experiment with $k=1, 3, 5, 10$.

For each prediction generated by T5, we also exploited its \texttt{score} function to assess the model's confidence on the provided input. 

The value returned by this function ranges from minus infinity to 0 and it is the log-likelihood ($ln$) of the prediction. Thus, if it is 0, it means that the likelihood of the prediction is 1 (\ie the maximum confidence, since $ln(1) = 0$), while when it goes towards minus infinity, the confidence tends to be 0. In our empirical study (\secref{sec:design}) we assess the reliability of the confidence level as a proxy for the quality of the predictions.

\section{Study Design} \label{sec:design}
The {\em goal} of our evaluation is to empirically assess the performance of the T5 model in code review automation tasks. The {\em context} consists of (i)  the datasets we presented in \secref{sec:t5}; and (ii) the dataset from our previous work \cite{Tufano:icse2021}. From now on we refer to our previously presented approach as the \emph{baseline}.
The study aims at tackling five research questions (RQs). 

\textbf{RQ$_1$: \emph{To what extent is T5 able to automatically recommend code changes to developers as reviewers would do?}} We provide as input to T5 a Java method $m_s$ submitted for review and assess the extent to which the model is able to provide as output a revised version of $m_s$ ($m_r$) implementing code changes that will be likely requested during the code review process. The idea here is that such a model could be used \emph{before} the code is submitted for review as an automated check for the contributor.

\textbf{RQ$_2$: \emph{To what extent is T5 able to automatically implement code changes recommended by reviewers?}} Given a Java method submitted for review ($m_s$) and a natural language comment ($c_{nl}$) in which a reviewer asks to implement specific code changes in $m_s$, we assess the ability of T5 to automatically revise $m_s$ to address $c_{nl}$ (thus obtaining a revised method $m_r$). \vspace{0.1cm}

The third RQ focuses on the novel code review-related task we introduce in this paper:

\textbf{RQ$_3$: \emph{To what extent is T5 able to automatically recommend changes in natural language as reviewers would do?}} In this RQ T5 is provided as input with a Java method submitted for review ($m_s$) and it is required to generate a natural language comment ($c_{nl}$) requesting code changes as reviewers would do. \vspace{0.1cm}

For RQ$_1$-RQ$_3$, we experiment with different variants of the T5 model. In particular, we assess the quality of T5 predictions for all three tasks when (i) the model is pre-trained or not; and (ii) the predictions have different confidence levels. Thanks to these analyses, we can answer our fourth RQ:

\textbf{RQ$_4$: \emph{What is the role played by the model pre-training on the performance of T5? How does the confidence of the predictions affects their quality?}} As explained in \secref{sub:hyper}, we perform an ablation study in which T5 is fine-tuned without any pre-training (\ie by starting from random weights in the neural network). This allows to assess the contribution of the pre-training to the performance of the model. As for the confidence of the predictions, we assess whether it can be used as a reliable proxy for the quality of the predictions (\ie the higher the confidence, the higher the likelihood the prediction is correct). If this is the case, such a finding would have implications for the usage of the T5 model in practice: A developer using the model could decide to receive recommendations having confidence higher than $t$, reducing the chances of receiving meaningless predictions. \vspace{0.1cm}

Finally, the last RQ compares the performance of the T5 model with that of the approach we presented in \cite{Tufano:icse2021}:

\textbf{RQ$_5$: \emph{What is the performance of T5 as compared to the state-of-the-art technique?}} We use the implementation and datasets from our previous work to compare the performance of the T5 model with the baseline \cite{Tufano:icse2021}.

\subsection{Data Collection and Analysis}
To answer the first four research questions, we experiment with the best configuration of both the pre-trained and non pre-trained T5 model on the test set of the fine-tuning dataset reported in \tabref{tab:datasets}. 

Remember that for each of the three tasks we support (\ie the ones that map to RQ$_1$, RQ$_2$, and RQ$_3$) the 16,779 test set instances are the same triplets $<m_s, c_{nl}, m_r>$. The only difference is that: in RQ$_1$ the model has been trained (and is tested) to take as input $m_s$ and produce $m_r$; in RQ$_2$ it takes as input $m_s$ and $c_{nl}$ and produces $m_r$; in RQ$_3$ it takes as input $m_s$ and produces $c_{nl}$. 

By running the models on the test sets, we report for each of the three tasks the percentage of ``perfect predictions'', namely the cases in which the output of the model is the expected one. For example, in the case of RQ$_3$, this means that the model was able, given $m_s$ as input, to generate a comment $c_{nl}$ identical to the one manually written by the reviewer who inspected $m_s$. 

Besides computing the perfect predictions, in RQ$_3$ (\ie the task in which the model is required to generate natural language text), we also compute the BLEU (Bilingual Evaluation Understudy) score of the predictions \cite{Papineni:2002}. BLEU assesses the quality of the automatically generated text. The BLEU score ranges between 0 and 1, with 1 indicating, in our case, that the natural language comment generated by the model is identical to the one manually written by the reviewer. We use the BLEU-4 variant, that computes the overlap in terms of 4-grams between the generated and the reference text.

In RQ$_1$ and RQ$_2$ (\ie in the tasks in which the model is required to generate code), we adopt instead the CodeBLEU \cite{ren2020codebleu}, a recently proposed similarity metric inspired by the BLEU score but tailored to assess the quality of automatically generated code. 

Differently from BLEU, CodeBLEU computes not only an ``n-gram based similarity'' but it also considers how similar the abstract syntax tree and the data-flow of the generated and the reference code are. Ren \etal \cite{ren2020codebleu}, who proposed the CodeBLEU, showed that their metric better correlates with developers' perception of code similarity as compared to the BLEU metric.

Concerning RQ$_4$, we compare the results (\ie perfect predictions, BLEU, CodeBLEU) achieved by the T5 model with and without pre-training. We also statistically compare the two models (\ie with/without pre-training) using the McNemar's test \cite{mcnemar} and Odds Ratios (ORs) on the perfect predictions they can generate. As for the confidence of the predictions, we take the best performing model (\ie the one with pre-training) and split its predictions into ten buckets based on their confidence $c$ going from 0.0 to 1.0 at steps of 0.1 (\ie the first interval includes all predictions having a confidence $c$ with 0 $< c \leq 0.1 $, the last interval has 0.9 $< c \leq 1$). Then, we report for each interval the percentage of perfect predictions.

Finally, in RQ$_5$, we compare T5 with the baseline \cite{Tufano:icse2021} on the two tasks automated in our previous work (\ie the ones related to our RQ$_1$ and RQ$_2$). 

As metrics for the comparisons, we used the percentage of perfect predictions and the CodeBLEU of the predictions. We compared the two techniques in several scenarios. First, we used the dataset from \cite{Tufano:icse2021} featuring 17,194 triplets $<m_s, c_{nl}, m_r>$. By performing some checks on this dataset, we noticed that a few instances (97) had comments ($c_{nl}$) not written in English or containing invalid unicode characters that did not allow our tokenizer to work. Thus, we excluded those instances from the training and the test sets shared by the authors. The training set has then been used to (i) train the baseline \cite{Tufano:icse2021}; and (ii) fine-tune the T5 model without any pre-training. In this way, we can compare the performance of the two models on the test set when trained on exactly the same data. Important to notice is that the baseline has been trained and tested on abstracted code (as done in \cite{Tufano:icse2021}), while T5 worked directly with the raw source code. 

On top of this, we also report the performance of the pre-trained T5 model when run on the test set from \cite{Tufano:icse2021}. This pre-trained model has been fine-tuned using the training dataset in \cite{Tufano:icse2021}. Clearly, this analysis favors T5 since it has been trained on more data (\ie the pre-training dataset). However, it provides additional hints into the role played by the pre-training and on the effectiveness of the T5 model in general. Besides reporting descriptive statistics, we statistically compare the two models using the McNemar's test \cite{mcnemar} and Odds Ratios (ORs) on the perfect predictions they can generate. Since multiple comparisons are involved (\eg comparing the pre-trained and the non pre-trained model to the baseline), we adjust the $p$-values using the Holm's correction \cite{Holm1979a}. 

\begin{figure*}
	\centering
	\includegraphics[width=0.8\linewidth]{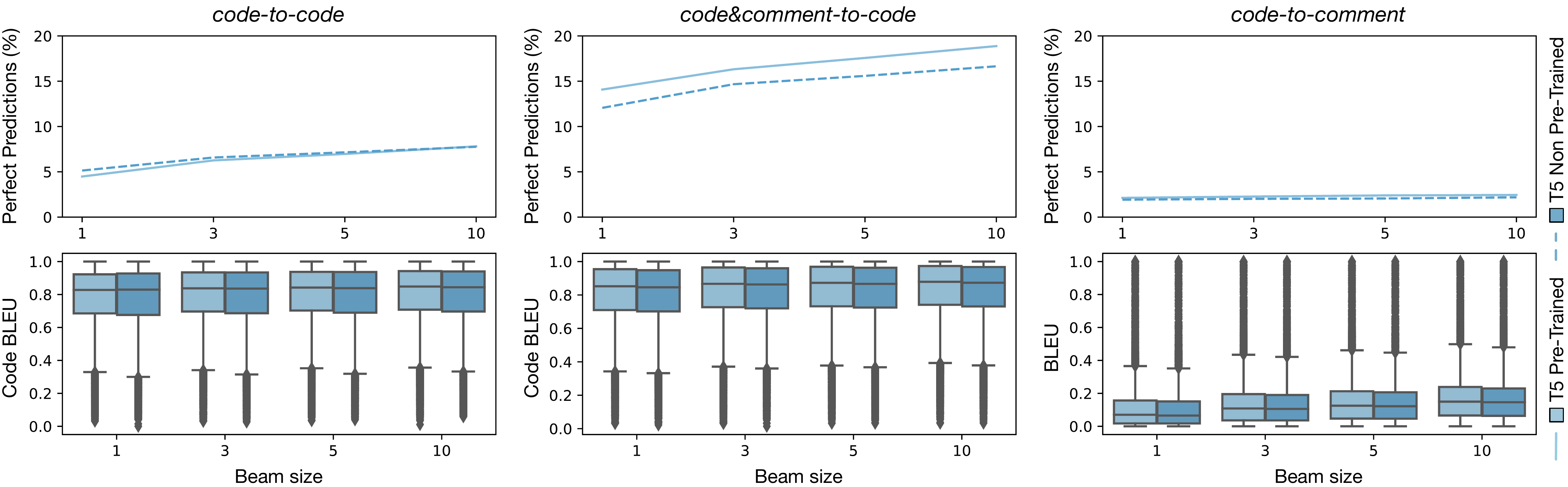}
	\caption{\vspace{-0.2cm}Results T5 dataset large}
	\label{fig:results_large}
\end{figure*}

\section{Results Discussion} \label{sec:results}
We start by answering \textbf{RQ$_1$-RQ$_3$} (\secref{sub:performance}), presenting the performance of T5 in the three tasks we aim at automating. Then, we discuss the impact on the performance of the pre-training and the reliability of the confidence level as a proxy for the quality of the predictions (\secref{sub:impact}). Finally, we compare T5 with the baseline \cite{Tufano:icse2021} (\secref{sub:comparison}).

\subsection{RQ$_1$-RQ$_3$: Performance of T5} \label{sub:performance}
\figref{fig:results_large} reports two graphs for each task. The line chart on top shows the percentage of perfect predictions ($y$-axis) achieved by T5 for different beam sizes ($x$-axis); the continuous line represents the pre-trained version of the model, while the dashed line the non pre-trained one. The boxplots at the bottom report the CodeBLEU for the two code-generation tasks (\ie \emph{code-to-code} and \emph{code\&comment-to-code}) and the BLEU score for the \emph{code-to-comment} task in which text is generated. Lighter blue represents the pre-trained model.

We start by commenting on the perfect predictions (line charts). At a first sight, the performance of the model might seem quite low. For example, in the case of \emph{code-to-code} at $k=1$ (\ie a single prediction is proposed by T5), both the pre-trained and the non pre-trained models achieve $\sim$5\% of perfect predictions (751 and 863 instances correctly predicted with and without pre-training, respectively). However, such a result should be considered in the context of what was reported by the state-of-the-art technique \cite{Tufano:icse2021} that, on a much simpler test dataset, achieved for the same task and same beam size 2.91\% of perfect predictions. 

Similar observations can be made for the \emph{code\&comment-to-code} task, where at $k=1$ T5 can generate 14.08\% (2,363 instances) and 12.06\% (2,024) perfect predictions when pre-trained and not, respectively. For this task, in our previous work \cite{Tufano:icse2021}, we achieved on a simpler dataset 12.16\% perfect predictions. We directly compare the two approaches in RQ$_5$. 

Interestingly, increasing the beam size from 1 to 10 does only result in marginal improvements for all tasks. The largest improvement is obtained for the \emph{code\&comment-to-code}, where we move from 14.08\% ($k=1$) to 18.88\% ($k=10$) of perfect predictions for the pre-trained model. Given the goal of our approach, we believe that the most relevant performance are those achieved at $k=1$. 

Indeed, providing several recommendations to inspect to a developer might be counterproductive, especially considering that the recommendations are entire methods in the case of the two code-generation tasks.

Moving to the \emph{code-to-comment} task, T5 struggles in formulating natural language comments identical to the ones written by reviewers. The pre-trained model, at $k=1$, generates 356 correct comments (2.12\%) against the 324 (1.93\%) of the non pre-trained model. These numbers only slightly increase at $k=10$, with a maximum of 2.44\% perfect predictions achieved with pre-training. 

\begin{figure*}
	\centering
	\includegraphics[width=\linewidth]{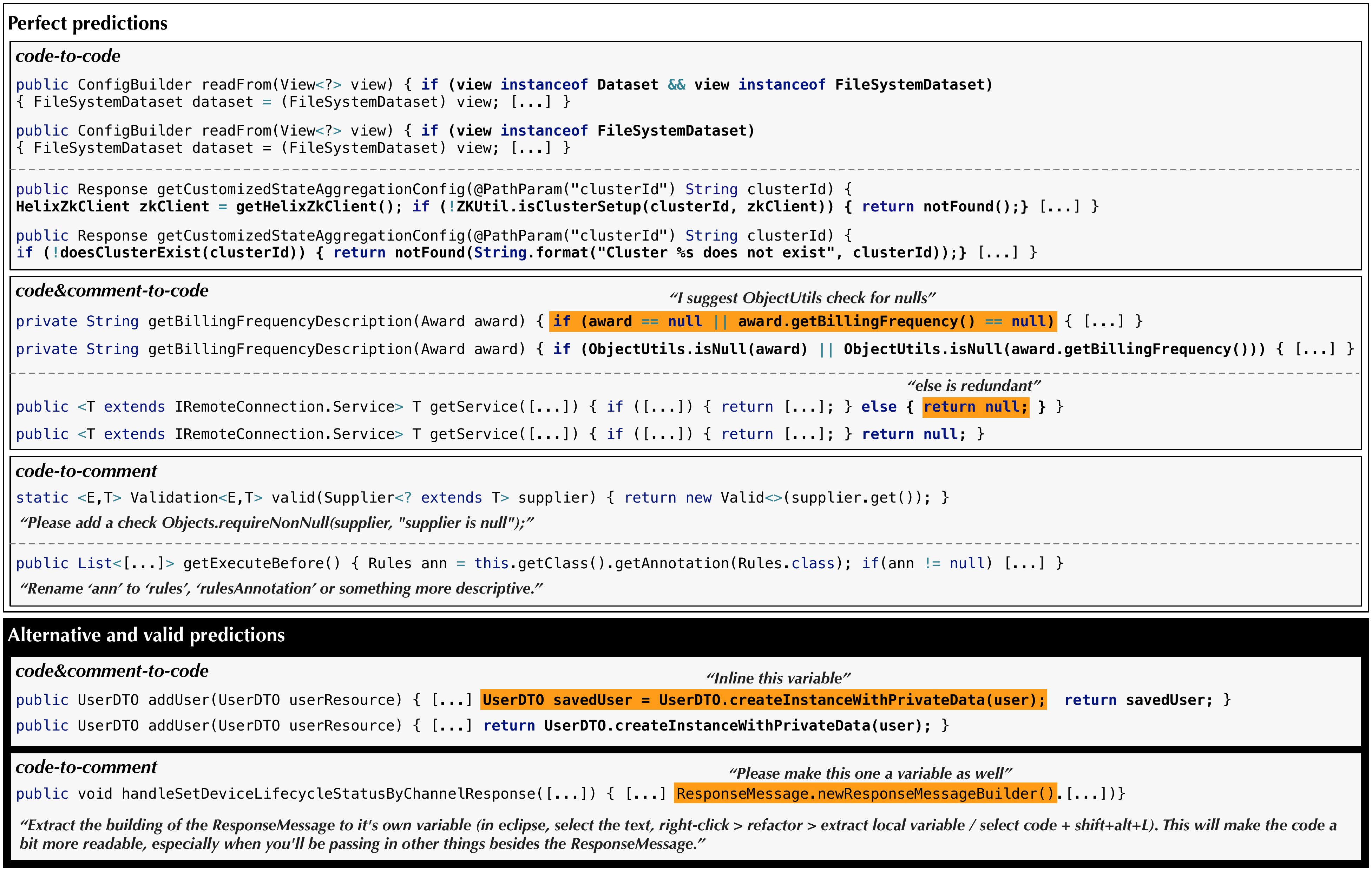}
	\caption{Examples of perfect and alternative predictions}
	\label{fig:qualitative}
\end{figure*}

The top part of \figref{fig:qualitative} shows two examples of perfect predictions generated by the model for each task. A dashed line separates the two examples within each task. For the \emph{code-to-code} task, the first code in each example represents the input of the model, while the second its output. We highlighted in bold the parts of code changed by the model and replaced irrelevant parts of the methods with [...] to save space. In the first \emph{code-to-code} example, T5 removes an unneeded \texttt{instanceof} check, since \texttt{FileSystemDataset} is a subclass of \texttt{Dataset}. Instead, the second example simplifies the checking for the existence of a cluster, providing a meaningful error message. This second case cannot be supported by the baseline \cite{Tufano:icse2021}, since it requires the introduction of new code tokens that were not present in the input code. Remember that,  these being perfect predictions, the implemented changes are identical to those performed by developers during code review.

For the \emph{code\&comment-to-code} task, the input provided by the model includes the comment written by the reviewer and requiring a specific change to the part of code highlighted in orange. In the first example, the reviewer suggests to use a specific object to perform the \texttt{null} check and T5 correctly implements the change. The second one is interesting because, despite the reviewer highlighting \texttt{return null} as the relevant code for their comment (``\emph{else is redundant}''), the model correctly understands that the action to take is the removal of the unneeded  \texttt{else} statement.

Finally, for the \emph{code-to-comment} task, we report the code provided as input to the model (first line) with the comment it generated as output (second line). In the first example, T5 suggests (as done by the real reviewer) to add a \texttt{null} check, also showing the code needed for its implementation. This code is not just a template, but it is suitable for the provided input code (it refers to the \texttt{supplier} object). In the second example, T5 suggests to rename an identifier, providing valid recommendations for the renaming.

Looking at the bottom of \figref{fig:results_large}, the results in terms of CodeBLEU show a median higher than 0.80 for all beam sizes and for both code-generation tasks.  However, while we report these values for completeness and for being consistent with what done in similar works \cite{Tufano:tosem2019,Watson:icse2020,Tufano:icse2021}, they say little about the quality of the predictions and they are mostly useful for future work that wants to compare with our approach (complete distributions are available in our replication package \cite{replication}). Indeed, it is difficult to properly interpret these values for two reasons. First, there is no accepted threshold above which good performance can be claimed. Second, as also done in previous works proposing models taking as input a code snippet and providing as output the same code ``revised'' in some way (\eg with a fixed bug \cite{Tufano:tosem2019}, with a single statement added \cite{Watson:icse2020}, or with review-related changes implemented \cite{Tufano:icse2021}), we computed the CodeBLEU between the predicted and the target code (two methods in our case). However, the input provided to the model is already quite similar to the target output, which means that a model taking as input a method and not implementing any change on it, is likely to obtain high values of CodeBLEU. For this reason, we mostly focus our discussion on perfect predictions. Concerning the BLEU score achieved in the \emph{code-to-comment} task, the median ranges around 0.10 (see \figref{fig:results_large}). Such a result is expected given the low percentage of perfect predictions achieved for this task.

Going back to the perfect predictions, the results reported in the line charts in \figref{fig:results_large} represent a lower bound for the performance of our approach. Indeed, we consider a prediction as ``perfect'' only if it is identical to the reference one. For example, in the case of the \emph{code-to-comment} task, the natural language comment generated by T5 is classified as correct only if it is equal to the reference one, including punctuation. 

However, it is possible that a natural language comment generated by T5 is different but semantically equivalent to the one written by the developer (\eg ``variable $v$ should be private'' \emph{vs} ``change $v$ visibility to private''). Similar observations hold for the two code-generation tasks (\eg a reviewer's comment could be addressed in different but semantically equivalent ways).

To have an idea on the number of valuable predictions present among those classified as ``wrong'' (\ie the non-perfect predictions), three authors manually analyzed a sample of 100 ``wrong'' predictions for each task (300 in total). The analysis was done in two meetings in which each instance was discussed by all three authors. The goal was to classify each instance into one of three categories: 
(i) ``semantically equivalent'' (\ie the generated code/comment is different but semantically equivalent to the reference one); (ii) ``alternative solution'' (\ie the generated code/comment is not semantically equivalent, but valuable); or (iii) ``wrong'' (\ie the generated code/comment is not meaningful for the provided input). Since we also computed the confidence for each of the predictions generated by T5, rather than randomly selecting the 300 instances to inspect, we decided to target for each task the top-100 wrong predictions generated by the model in terms of confidence. Indeed, those cases are particularly interesting, since they represent wrong predictions for which, however, the model is quite confident. 

\begin{table}[ht]
	\centering
	\scriptsize
	\caption{Manual analysis of 100 ``wrong'' predictions per task}
        \label{tab:manual}
	\begin{tabular}{lrrr}
		\toprule
		\textbf{Task} & \textbf{Semantically Equivalent} & \textbf{Alternative Solution} & \textbf{Wrong} \\ 
		\midrule
		\emph{code-to-code} & 1 & 10 & 89 \\
		\emph{code\&comment-to-code} & 6 & 56 & 38\\
		\emph{code-to-comment} & 36 & 10 & 54\\
		\bottomrule
	\end{tabular}
\end{table}

\tabref{tab:manual} shows the results of our manual analysis. For the \emph{code-to-code} we observed that, in most cases (89\%) the model actually generates wrong predictions that are not inline with the changes implemented by the developer. There are few exceptions to these cases, mostly related to small changes in which the model made a decision different from that one of the developer but still valid (\eg extracting a string into a variable and using a different name for the extracted variable). More interesting are the results for the other two tasks. 

In the case of \emph{code\&comment-to-code}, we found that 62 out of the 100 ``wrong'' predictions we inspected were actually valid implementations of the change recommended by the reviewer. One example is presented at the bottom of \figref{fig:qualitative} (black background), where we show the input provided to the model (\ie the code in the first line and the reviewer's comment ``\emph{Inline this variable}'') and the output of the model right below. T5 successfully addressed the reviewer's comment. 

However, the prediction is different from the target implementation, since the latter also includes another change that was not explicitly required in the code review. This case is representative of all 56 instances we classified as ``alternative solutions'' for this task and, given the goal of the \emph{code\&comment-to-code}, we believe they represent good predictions.

Finally, also for the \emph{code-to-comment} task, we found a large number of ``wrong'' predictions that are actually valuable, with 36 of them even being semantically equivalent (\ie T5 formulated a comment asking the same changes required by the reviewer, but using a different wording). One example is reported at the very bottom of \figref{fig:qualitative}. While the model only received the code as input we also show the original reviewer's comment (\ie ``\emph{Please make this one a variable as well}'') to make it easier to assess the relevance of the comment generated by T5 (\ie ``\emph{Extract the building ...}''). 

Overall, our analysis showed that the perfect predictions really represent a lower bound for the performance of T5, especially for the two tasks in which natural language comments are involved.

\begin{figure}
	\centering
	\includegraphics[width=\linewidth]{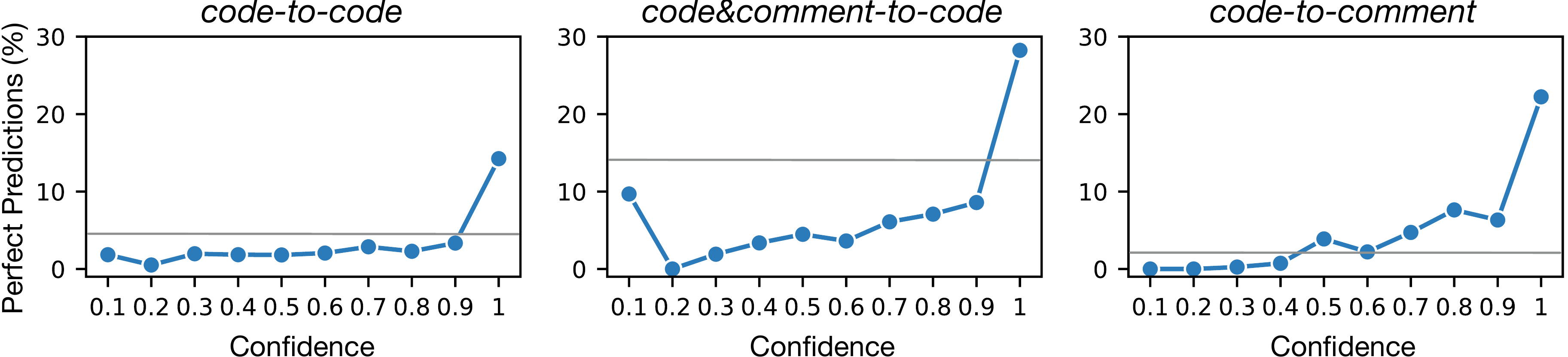}
	\caption{\vspace{-0.3cm}Perfect predictions by confidence of the model}
	\label{fig:confidence}
\end{figure}

\subsection{RQ$_4$: Pre-training and confidence}\label{sub:impact}
In \figref{fig:results_large} we observed better performance for the pre-trained model in the \emph{code\&comment-to-code} and in the \emph{code-to-comment} task, while the non pre-trained model performed better in the \emph{code-to-code} task. The results of the McNemar's test on the predictions at $k$=1, confirm such findings: besides the significant difference confirmed for all tasks ($p$-value $<$ 0.01), the ORs indicate 85\% and 59\% higher odds of obtaining a perfect prediction using the pre-trained model in the \emph{code\&comment-to-code} (OR=1.85) and in the \emph{code-to-comment} (OR=1.59) task, while odds are 34\% lower in the \emph{code-to-code} task (OR=0.66). Two observations are worth to be made. First, overall, the pre-trained model seems to represent a more valuable solution. Second, the lack of improvement in the \emph{code-to-code} task can be explained by the pre-training and fine-tuning we performed. Indeed, the \emph{code-to-code} task only focuses on source code, with no natural language in the input nor in the output. The fine-tuning stage, focused on source code, was probably sufficient to the model to learn about the code syntax and the possible transformations to perform. The additional pre-training, also including technical English, did not benefit the model for the \emph{code-to-code} task. The other two tasks, instead, either include natural language as input (\emph{code\&comment-to-code}) or require its generation as output (\emph{code-to-comment}), obtaining a boost of performance from the pre-training.

\figref{fig:confidence} depicts the percentage of perfect predictions ($y$-axis) within each confidence interval (from  0.0-0.1 up to 0.9-1.0, $x$-axis) when using the pre-trained model and $k$=1. To better interpret the reported results, the gray line represents the overall performance of the model when considering all predictions (\eg 4.48\% of perfect predictions for the \emph{code-to-code} task). 

In all three tasks, we observe a clear trend, with the predictions in the highest confidence bucket (0.9-1.0) ensuring substantially better performance than the overall trend. When only considering the predictions in this bucket, the percentage of perfect predictions increases to: 14.24\% for \emph{code-to-code} (from an overall 4.48\%), 28.23\% for \emph{code\&comment-to-code} (overall=14.08\%), and 22.23\% for \emph{code-to-comment} (overall=2.12\%). Considering the complexity of the addressed tasks, the jump in performance is substantial and indicates the usability of the confidence level as a proxy for the prediction quality. Also, while the percentage of perfect predictions is quite limited, with seven out of ten predictions being wrong in the best-case scenario (28.23\% for \emph{code\&comment-to-code}), it is worth considering what previously observed in our manual analysis, with ``valuable'' predictions which are classified as ``wrong'' in our quantitative analysis.

\subsection{RQ$_5$: Comparison with the baseline \cite{Tufano:icse2021}}\label{sub:comparison}
\begin{figure}
	\centering
	\includegraphics[width=\linewidth]{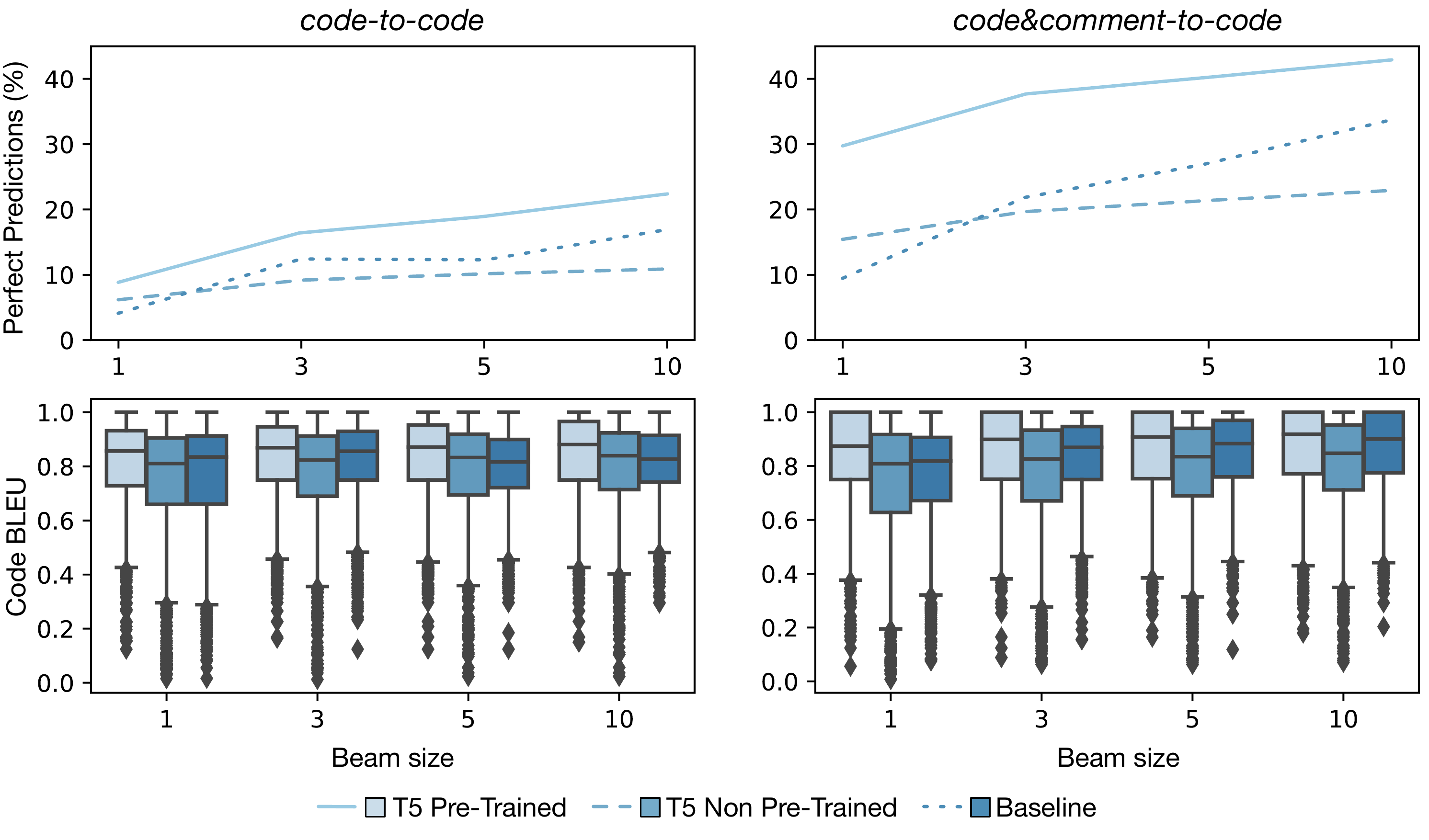}
	\caption{\vspace{-0.3cm}T5 \emph{vs.} baseline \cite{Tufano:icse2021}}
	\label{fig:results_small}
\end{figure}

\figref{fig:results_small} compares the performance achieved by the T5 model with those obtained by the baseline \cite{Tufano:icse2021}. 

In the line charts the continuous lines represent the pre-trained T5, the dashed lines non pre-trained T5, and the dotted lines the baseline. Two important points are worth remembering: First, the results in \figref{fig:results_small} have been computed on the test set used in \cite{Tufano:icse2021}. Indeed, the performance in terms of perfect predictions are substantially higher as compared to those in \figref{fig:results_large} (see values on the $y$-axis), due to the simpler instances featured in this dataset. Second, the baseline has been trained and tested on abstracted code (as in the original paper), while T5 worked on raw source code.

When $k$=1, T5 achieves substantially better performance. The results of the statistical test in \tabref{tab:stats-comparison} always show a significant difference in favor of T5 (adjusted $p$-value $<$ 0.01), with ORs ranging from 1.69 (non pre-trained T5 \emph{vs} \cite{Tufano:icse2021} in the \emph{code-to-code} task) to 11.48 (pre-trained T5 \emph{vs} \cite{Tufano:icse2021} in the \emph{code\&comment-to-code} task). The pre-trained T5 in this case performs better than the non pre-trained one for both tasks. This is likely due to the limited size of the fine-tuning dataset used in this comparison. Indeed, to have a fair comparison with \cite{Tufano:icse2021}, we fine-tuned T5 on the training set we used in \cite{Tufano:icse2021} and composed by $\sim$13.5k instances (\emph{vs} the $\sim$134k we had in our fine-tuning dataset when answering RQ$_1$-RQ$_4$). This is probably not sufficient to effectively train a large model such as T5, and makes the instances used in the pre-training fundamental to further learn about the language. Still, even without pre-training, T5 outperforms the baseline when $k$=1. For example, in the \emph{code\&comment-to-code} task, the baseline achieves 9.48\% perfect predictions, against the 15.46\% of the non pre-trained T5, and the 29.74\% of the pre-trained T5. The baseline observes a stronger improvement with the increasing of $k$ (\ie the beam size) as compared to T5 (see \figref{fig:results_small}). We believe this is due to usage of the abstraction. Indeed, when working with abstracted code the ``search space'' (\ie the number of possible solutions that can be generated with the given vocabulary) is much more limited since the model does not deal with identifiers and literals. Attempting ten predictions in a smaller search space is more likely to result in correct predictions. The results of the CodeBLEU confirm the trend observed with the perfect predictions, with the pre-trained T5 being the best model.

We also looked at the union of perfect predictions generated by the two approaches on the test set to verify the complementarity of the techniques. On the \textit{code-to-code} (\textit{code\&comment-to-code}) task we observed that 15\% (24\%) of perfect predictions are shared by both approaches (\ie both succeed), 65\% (70\%) are perfect predictions only for T5, and 20\% (6\%) only for the baseline.

\begin{table}[ht]
	\centering
	\scriptsize
	\caption{RQ$_5$: McNemar's test (adj. $p$-value and OR)}
	\vspace{-0.3cm}
        \label{tab:stats-comparison}
	\begin{tabular}{llrr}
		\toprule
		\textbf{Task} & \textbf{Test} & \textbf{\emph{p}-value} & \textbf{OR} \\ 
		\midrule
		\multirow{3}{*}{\emph{code-to-code}} & T5 pre-trained \emph{vs} \cite{Tufano:icse2021} & $<$0.01 & 2.90\\
		& T5 non pre-trained \emph{vs} \cite{Tufano:icse2021}  & $<$0.01 &1.69\\
		& T5 pre-trained \emph{vs} T5 non pre-trained  & $<$0.01 &2.50\\\midrule
		
		\multirow{3}{*}{\emph{code\&comment-to-code}} & T5 pre-trained \emph{vs} \cite{Tufano:icse2021} & $<$0.01 &11.48\\
		&  T5 non pre-trained \emph{vs} \cite{Tufano:icse2021} & $<$0.01 & 2.38\\
		&  T5 pre-trained \emph{vs} T5 non pre-trained & $<$0.01 & 5.69\\

		\bottomrule
	\end{tabular}
	\vspace{-0.3cm}
\end{table}
\section{Threats to Validity} \label{sec:threats}

\textbf{Construct validity.} As explained in \secref{sec:t5} we took care of cleaning the datasets used in our study by removing duplicates and noisy data points to the extent possible. Still, we are aware that problematic instances may be present, especially in the new (large) dataset we built. This manifests, for example, in non-English comments, or in some wrong ``links'' between comments and implementation (\eg we assume that $m_r$ implemented a change described in $c_nl$ while, in fact, it implemented another change).

\textbf{Internal validity.} We did not fully explore the role played by the T5 parameters on its performance. Indeed, our hyperparameter tuning was limited to variations in the learning rate, as done in previous work \cite{mastropaolo2021studying}. For the other parameters we relied on the best architecture identified by Raffel \etal \cite{raffel2019exploring}. We acknowledge that additional tuning can result in improved performance.

\textbf{External validity.} RQ$_1$-RQ$_4$ have been answered using a dataset being one order of magnitude larger as compared to our previous work on automating code review tasks \cite{Tufano:icse2021}. However, our findings are limited to Java. Concerning RQ$_5$ in which we compare with the baseline \cite{Tufano:icse2021}, we only used the dataset presented in \cite{Tufano:icse2021}. This is due to the fact that our previous approach \cite{Tufano:icse2021} requires code abstraction and, as previously explained, cannot work on instances having new identifiers and literals inserted during the code review process. The new dataset used in this paper has not been built with such a constraint in mind and, thus, it is not suitable for direct comparison.

\section{Related Work} \label{sec:related}

Our work relates to three research areas: (i) DL techniques to automate software-related tasks, (ii) empirical studies on code review, and (iii) works providing recommendations on how to optimize the code review process and/or presenting techniques to partially automate it. Here we focus on the third research area, while for the first two we point the reader to the systematic literature reviews by Watson \etal~\cite{watson2020systematic} (deep learning in software engineering) and by Davila and Nunes \cite{davila2021systematic} (modern code review).

\textbf{Optimizing/automating the code review process.} By studying tools and techniques supporting code review, Tymchuk \etal \cite{tymchuk2015code} concluded that popular code review platforms (\eg Gerrit, Code Flow, Phabricator) mostly offer the same basic functionalities with little support for automating tasks. Such a finding has been confirmed by Pascarella \etal~\cite{pascarella2018information}. Also, in a study performed by Lewis~\etal~\cite{lewis2013does} at Google, the authors show that while developers are excited by the idea of embracing automated solutions for code review, they find current solutions not to be ready for daily use. Starting from these observations, researchers studied possible optimizations of the review process: Baum \etal~\cite{baum2017optimal} investigate the effect of ordering submitted changes in alternative ways rather than in alphabetical order that, as shown by Barnett \etal~\cite{Barnett:icse2015} and Baum and Schneider~\cite{baum2016need}, is sub-optimal. Baum \etal~\cite{baum2017optimal} concluded that smarter ordering is needed as the size of the patch increases, and suggest to aggregate changed parts by relatedness. 

Di Biase~\etal~\cite{di2019effects} studied the impact of the patch size on the review's effectiveness, finding that smaller patches, while not increasing the defects found, affect how reviewers approach their task. Spadini \etal~\cite{spadini2019test} compared the effectiveness of a standard code review process with test-driven code review (TDR), \ie the reviewer inspects the changed test code before the production code. They show that TDR does not boost the code review effectiveness. 

Several researchers~\cite{hoang2019deepjit,pascarella2019fine,Wattanakriengkrai2020linelevel} suggest exploiting defect prediction models during code review. Similarly, Balachandran~\cite{Balachandran2013} and Singh \etal~\cite{singh2017evaluating} suggest the use of static analysis tools to automatically spot coding standard violations and common defects. 

Concerning the automation of specific code review tasks, authors proposed techniques to optimize the reviewers' assignment. For example, Al-Zubaidi \etal~\cite{al2020workload} in open source and Chouchen \etal~\cite{chouchen2021whoreview} in industrial contexts show how a multi-objective search-based approach can simplify the code review triaging process.

Shi \etal~\cite{shi2019automatic} and Chouchen \etal~\cite{chouchen2021whoreview} look at the automation of code review from a similar perspective. Shi \etal~\cite{shi2019automatic} present a DL model taking as input the code submitted for review and the revised code implementing the changes recommended by reviewers and providing as output whether the change can be accepted or not. Note that the change(s) required by the reviewer(s) are not considered by the model. Chouchen \etal~\cite{chouchen2021whoreview} use instead a set of quality metrics as features for machine learning algorithms to classify the quality of the code submitted for review. Recently, Hellendoorn \etal \cite{hellendoorn2021} focus on the prediction of the location of a possible reviewer's comment, showing that even this simple task is challenging to automate. 

The above discussed techniques \cite{shi2019automatic,chouchen2021whoreview,hellendoorn2021} are complementary to the approach we presented in \cite{Tufano:icse2021} (and, as a consequence, to the models experimented in this work). 

While Shi \etal~\cite{shi2019automatic} and Chouchen \etal~\cite{chouchen2021whoreview} assess the code under review through a ``boolean answer'' (\ie accepted/rejected or well-written/badly-written), we attempt the automation of code changes implemented in code review. Also, the approach by Hellendoorn \etal could be combined with the automation of the \textit{code-to-comment} task we presented.
\section{Conclusion and Future Work} \label{sec:conclusion}
Our paper starts by discussing limitations in the approach we recently proposed to automate code review tasks \cite{Tufano:icse2021}. We highlighted that the usage of code abstraction does not allow to support non-trivial code review scenarios requiring code changes resulting in the introduction of new identifiers/literals. Hence, we proposed the usage of a pre-trained T5 model \cite{raffel2019exploring} relying on a SentencePiece \cite{Kudo:sentencePiece} tokenizer to overcome such a limitation and work directly on raw source code. Our empirical evaluation, performed on a much larger and realistic code review dataset, shows the improvements brought by the T5 model that represents a step forward as compared to the state-of-the-art \cite{Tufano:icse2021} both in terms of applicability (\ie scenarios in which it can be applied) and performance. Still, the level of actual performance observed makes these techniques far from being deployable in practice, calling for more research in code review automation.

Our future research agenda will be focused on designing improved solutions to boost the prediction accuracy of these techniques (\eg by combining different representations of code \cite{chakraborty2021multimodal} and/or by exploiting the model's confidence as a possible filter to select only high-quality recommendations). 

The code and data used in our study are publicly available \cite{replication}. 

\section*{Acknowledgment}
This project has received funding from the European Research Council (ERC) under the European Union's Horizon 2020 research and innovation programme (grant agreement No. 851720). W\&M has been supported in part by the NSF CCF-1955853 and CCF-2007246 grants. Any opinions, findings, and conclusions expressed herein are the authors' and do not necessarily reflect those of the sponsors. 

\bibliographystyle{ACM-Reference-Format}
\bibliography{main}


\begin{thebibliography}{51}


\ifx \showCODEN    \undefined \def \showCODEN     #1{\unskip}     \fi
\ifx \showDOI      \undefined \def \showDOI       #1{#1}\fi
\ifx \showISBNx    \undefined \def \showISBNx     #1{\unskip}     \fi
\ifx \showISBNxiii \undefined \def \showISBNxiii  #1{\unskip}     \fi
\ifx \showISSN     \undefined \def \showISSN      #1{\unskip}     \fi
\ifx \showLCCN     \undefined \def \showLCCN      #1{\unskip}     \fi
\ifx \shownote     \undefined \def \shownote      #1{#1}          \fi
\ifx \showarticletitle \undefined \def \showarticletitle #1{#1}   \fi
\ifx \showURL      \undefined \def \showURL       {\relax}        \fi
\providecommand\bibfield[2]{#2}
\providecommand\bibinfo[2]{#2}
\providecommand\natexlab[1]{#1}
\providecommand\showeprint[2][]{arXiv:#2}

\bibitem[\protect\citeauthoryear{??}{Ger}{[n.d.]}]%
        {Gerrit}
 \bibinfo{year}{[n.d.]}\natexlab{}.
\newblock \bibinfo{title}{Gerrit. \url{https://www.gerritcodereview.com/}}.
\newblock
\newblock


\bibitem[\protect\citeauthoryear{??}{Git}{[n.d.]}]%
        {GitHub}
 \bibinfo{year}{[n.d.]}\natexlab{}.
\newblock \bibinfo{title}{GitHub. \url{https://github.com/}}.
\newblock
\newblock


\bibitem[\protect\citeauthoryear{??}{lan}{[n.d.]}]%
        {langdetect}
 \bibinfo{year}{[n.d.]}\natexlab{}.
\newblock \bibinfo{title}{langdetect}.
\newblock \bibinfo{howpublished}{\url{https://pypi.org/project/langdetect/}}.
\newblock


\bibitem[\protect\citeauthoryear{??}{Liz}{[n.d.]}]%
        {Lizard}
 \bibinfo{year}{[n.d.]}\natexlab{}.
\newblock \bibinfo{title}{Lizard. \url{https://github.com/terryyin/lizard/}}.
\newblock
\newblock


\bibitem[\protect\citeauthoryear{??}{ghs}{[n.d.]}]%
        {ghs}
 \bibinfo{year}{[n.d.]}\natexlab{}.
\newblock \bibinfo{title}{MSR mining platform}.
\newblock \bibinfo{howpublished}{\url{https://seart-ghs.si.usi.ch}}.
\newblock


\bibitem[\protect\citeauthoryear{??}{pyc}{[n.d.]}]%
        {pycld3}
 \bibinfo{year}{[n.d.]}\natexlab{}.
\newblock \bibinfo{title}{pycld3}.
\newblock \bibinfo{howpublished}{\url{https://pypi.org/project/pycld3/}}.
\newblock


\bibitem[\protect\citeauthoryear{??}{so_}{[n.d.]}]%
        {so_dump}
 \bibinfo{year}{[n.d.]}\natexlab{}.
\newblock \bibinfo{title}{Stack Exchange Dumps}.
\newblock
  \bibinfo{howpublished}{\url{https://archive.org/details/stackexchange}}.
\newblock


\bibitem[\protect\citeauthoryear{??}{rep}{2021}]%
        {replication}
 \bibinfo{year}{2021}\natexlab{}.
\newblock \bibinfo{title}{{Replication Package}}.
\newblock
  \bibinfo{howpublished}{{\url{https://github.com/RosaliaTufano/code_review_automation}}}.
\newblock


\bibitem[\protect\citeauthoryear{Al-Zubaidi, Thongtanunam, Dam,
  Tantithamthavorn, and Ghose}{Al-Zubaidi et~al\mbox{.}}{2020}]%
        {al2020workload}
\bibfield{author}{\bibinfo{person}{Wisam Haitham~Abbood Al-Zubaidi},
  \bibinfo{person}{Patanamon Thongtanunam}, \bibinfo{person}{Hoa~Khanh Dam},
  \bibinfo{person}{Chakkrit Tantithamthavorn}, {and} \bibinfo{person}{Aditya
  Ghose}.} \bibinfo{year}{2020}\natexlab{}.
\newblock \showarticletitle{Workload-aware reviewer recommendation using a
  multi-objective search-based approach}. In
  \bibinfo{booktitle}{\emph{Proceedings of the 16th ACM International
  Conference on Predictive Models and Data Analytics in Software Engineering}}.
  \bibinfo{pages}{21--30}.
\newblock


\bibitem[\protect\citeauthoryear{Bacchelli and Bird}{Bacchelli and
  Bird}{2013}]%
        {bacchelli2013expectations}
\bibfield{author}{\bibinfo{person}{Alberto Bacchelli} {and}
  \bibinfo{person}{Christian Bird}.} \bibinfo{year}{2013}\natexlab{}.
\newblock \showarticletitle{Expectations, outcomes, and challenges of modern
  code review}. In \bibinfo{booktitle}{\emph{Proceedings of the 2013
  international conference on software engineering}}. IEEE Press,
  \bibinfo{pages}{712--721}.
\newblock


\bibitem[\protect\citeauthoryear{Balachandran}{Balachandran}{2013}]%
        {Balachandran2013}
\bibfield{author}{\bibinfo{person}{Vipin Balachandran}.}
  \bibinfo{year}{2013}\natexlab{}.
\newblock \showarticletitle{Reducing human effort and improving quality in peer
  code reviews using automatic static analysis and reviewer recommendation}. In
  \bibinfo{booktitle}{\emph{2013 35th International Conference on Software
  Engineering (ICSE)}}. \bibinfo{pages}{931--940}.
\newblock
\urldef\tempurl%
\url{https://doi.org/10.1109/ICSE.2013.6606642}
\showDOI{\tempurl}


\bibitem[\protect\citeauthoryear{Barnett, Bird, Brunet, and Lahiri}{Barnett
  et~al\mbox{.}}{2015}]%
        {Barnett:icse2015}
\bibfield{author}{\bibinfo{person}{Mike Barnett}, \bibinfo{person}{Christian
  Bird}, \bibinfo{person}{Jo\~{a}o Brunet}, {and} \bibinfo{person}{Shuvendu~K.
  Lahiri}.} \bibinfo{year}{2015}\natexlab{}.
\newblock \showarticletitle{Helping Developers Help Themselves: Automatic
  Decomposition of Code Review Changesets}. In
  \bibinfo{booktitle}{\emph{Proceedings of the 37th International Conference on
  Software Engineering - Volume 1}} \emph{(\bibinfo{series}{ICSE '15})}.
  \bibinfo{pages}{134--144}.
\newblock


\bibitem[\protect\citeauthoryear{Baum and Schneider}{Baum and
  Schneider}{2016}]%
        {baum2016need}
\bibfield{author}{\bibinfo{person}{Tobias Baum} {and} \bibinfo{person}{Kurt
  Schneider}.} \bibinfo{year}{2016}\natexlab{}.
\newblock \showarticletitle{On the need for a new generation of code review
  tools}. In \bibinfo{booktitle}{\emph{International Conference on
  Product-Focused Software Process Improvement}}. Springer,
  \bibinfo{pages}{301--308}.
\newblock


\bibitem[\protect\citeauthoryear{Baum, Schneider, and Bacchelli}{Baum
  et~al\mbox{.}}{2017}]%
        {baum2017optimal}
\bibfield{author}{\bibinfo{person}{Tobias Baum}, \bibinfo{person}{Kurt
  Schneider}, {and} \bibinfo{person}{Alberto Bacchelli}.}
  \bibinfo{year}{2017}\natexlab{}.
\newblock \showarticletitle{On the optimal order of reading source code changes
  for review}. In \bibinfo{booktitle}{\emph{2017 IEEE International Conference
  on Software Maintenance and Evolution (ICSME)}}. IEEE,
  \bibinfo{pages}{329--340}.
\newblock


\bibitem[\protect\citeauthoryear{Bavota and Russo}{Bavota and Russo}{2015}]%
        {Bavota:icsme2015}
\bibfield{author}{\bibinfo{person}{Gabriele Bavota} {and}
  \bibinfo{person}{Barbara Russo}.} \bibinfo{year}{2015}\natexlab{}.
\newblock \showarticletitle{Four eyes are better than two: On the impact of
  code reviews on software quality}. In \bibinfo{booktitle}{\emph{{IEEE}
  International Conference on Software Maintenance and Evolution, ({ICSME})}}.
  \bibinfo{pages}{81--90}.
\newblock


\bibitem[\protect\citeauthoryear{Bosu and Carver}{Bosu and Carver}{2013}]%
        {Bosu:2013}
\bibfield{author}{\bibinfo{person}{A. Bosu} {and} \bibinfo{person}{J.~C.
  Carver}.} \bibinfo{year}{2013}\natexlab{}.
\newblock \showarticletitle{Impact of Peer Code Review on Peer Impression
  Formation: A Survey}. In \bibinfo{booktitle}{\emph{2013 ACM / IEEE
  International Symposium on Empirical Software Engineering and Measurement}}.
  \bibinfo{pages}{133--142}.
\newblock


\bibitem[\protect\citeauthoryear{Chakraborty and Ray}{Chakraborty and
  Ray}{2021}]%
        {chakraborty2021multimodal}
\bibfield{author}{\bibinfo{person}{Saikat Chakraborty} {and}
  \bibinfo{person}{Baishakhi Ray}.} \bibinfo{year}{2021}\natexlab{}.
\newblock \bibinfo{title}{On Multi-Modal Learning of Editing Source Code}.
\newblock
\newblock
\showeprint[arxiv]{2108.06645}~[cs.SE]


\bibitem[\protect\citeauthoryear{Chouchen, Ouni, Mkaouer, Kula, and
  Inoue}{Chouchen et~al\mbox{.}}{2021}]%
        {chouchen2021whoreview}
\bibfield{author}{\bibinfo{person}{Moataz Chouchen}, \bibinfo{person}{Ali
  Ouni}, \bibinfo{person}{Mohamed~Wiem Mkaouer},
  \bibinfo{person}{Raula~Gaikovina Kula}, {and} \bibinfo{person}{Katsuro
  Inoue}.} \bibinfo{year}{2021}\natexlab{}.
\newblock \showarticletitle{WhoReview: A multi-objective search-based approach
  for code reviewers recommendation in modern code review}.
\newblock \bibinfo{journal}{\emph{Applied Soft Computing}}
  \bibinfo{volume}{100} (\bibinfo{year}{2021}), \bibinfo{pages}{106908}.
\newblock


\bibitem[\protect\citeauthoryear{Dabic, Aghajani, and Bavota}{Dabic
  et~al\mbox{.}}{2021}]%
        {dabic2021sampling}
\bibfield{author}{\bibinfo{person}{Ozren Dabic}, \bibinfo{person}{Emad
  Aghajani}, {and} \bibinfo{person}{Gabriele Bavota}.}
  \bibinfo{year}{2021}\natexlab{}.
\newblock \showarticletitle{Sampling Projects in GitHub for {MSR} Studies}. In
  \bibinfo{booktitle}{\emph{18th {IEEE/ACM} International Conference on Mining
  Software Repositories, {MSR} 2021}}. \bibinfo{publisher}{{IEEE}},
  \bibinfo{pages}{560--564}.
\newblock


\bibitem[\protect\citeauthoryear{Davila and Nunes}{Davila and Nunes}{2021}]%
        {davila2021systematic}
\bibfield{author}{\bibinfo{person}{Nicole Davila} {and} \bibinfo{person}{Ingrid
  Nunes}.} \bibinfo{year}{2021}\natexlab{}.
\newblock \showarticletitle{A systematic literature review and taxonomy of
  modern code review}.
\newblock \bibinfo{journal}{\emph{Journal of Systems and Software}}
  (\bibinfo{year}{2021}), \bibinfo{pages}{110951}.
\newblock


\bibitem[\protect\citeauthoryear{di~Biase, Bruntink, van Deursen, and
  Bacchelli}{di~Biase et~al\mbox{.}}{2019}]%
        {di2019effects}
\bibfield{author}{\bibinfo{person}{Marco di Biase}, \bibinfo{person}{Magiel
  Bruntink}, \bibinfo{person}{Arie van Deursen}, {and} \bibinfo{person}{Alberto
  Bacchelli}.} \bibinfo{year}{2019}\natexlab{}.
\newblock \showarticletitle{The effects of change decomposition on code
  review—a controlled experiment}.
\newblock \bibinfo{journal}{\emph{PeerJ Computer Science}}  \bibinfo{volume}{5}
  (\bibinfo{year}{2019}), \bibinfo{pages}{e193}.
\newblock


\bibitem[\protect\citeauthoryear{Hellendoorn, Tsay, Mukherjee, and
  Hirzel}{Hellendoorn et~al\mbox{.}}{2021}]%
        {hellendoorn2021}
\bibfield{author}{\bibinfo{person}{Vincent~J Hellendoorn},
  \bibinfo{person}{Jason Tsay}, \bibinfo{person}{Manisha Mukherjee}, {and}
  \bibinfo{person}{Martin Hirzel}.} \bibinfo{year}{2021}\natexlab{}.
\newblock \showarticletitle{Towards automating code review at scale}. In
  \bibinfo{booktitle}{\emph{Proceedings of the 29th ACM Joint Meeting on
  European Software Engineering Conference and Symposium on the Foundations of
  Software Engineering}}. \bibinfo{pages}{1479--1482}.
\newblock


\bibitem[\protect\citeauthoryear{Hoang, Dam, Kamei, Lo, and Ubayashi}{Hoang
  et~al\mbox{.}}{2019}]%
        {hoang2019deepjit}
\bibfield{author}{\bibinfo{person}{Thong Hoang}, \bibinfo{person}{Hoa~Khanh
  Dam}, \bibinfo{person}{Yasutaka Kamei}, \bibinfo{person}{David Lo}, {and}
  \bibinfo{person}{Naoyasu Ubayashi}.} \bibinfo{year}{2019}\natexlab{}.
\newblock \showarticletitle{DeepJIT: an end-to-end deep learning framework for
  just-in-time defect prediction}. In \bibinfo{booktitle}{\emph{2019 IEEE/ACM
  16th International Conference on Mining Software Repositories (MSR)}}. IEEE,
  \bibinfo{pages}{34--45}.
\newblock


\bibitem[\protect\citeauthoryear{Holm}{Holm}{1979}]%
        {Holm1979a}
\bibfield{author}{\bibinfo{person}{Sture Holm}.}
  \bibinfo{year}{1979}\natexlab{}.
\newblock \showarticletitle{A simple sequentially rejective multiple test
  procedure}.
\newblock \bibinfo{journal}{\emph{Scandinavian journal of statistics}}
  (\bibinfo{year}{1979}), \bibinfo{pages}{65--70}.
\newblock


\bibitem[\protect\citeauthoryear{Husain, Wu, Gazit, Allamanis, and
  Brockschmidt}{Husain et~al\mbox{.}}{2019}]%
        {Java:CodeSearchNet}
\bibfield{author}{\bibinfo{person}{Hamel Husain}, \bibinfo{person}{Ho{-}Hsiang
  Wu}, \bibinfo{person}{Tiferet Gazit}, \bibinfo{person}{Miltiadis Allamanis},
  {and} \bibinfo{person}{Marc Brockschmidt}.} \bibinfo{year}{2019}\natexlab{}.
\newblock \showarticletitle{CodeSearchNet Challenge: Evaluating the State of
  Semantic Code Search}.
\newblock \bibinfo{journal}{\emph{CoRR}}  \bibinfo{volume}{abs/1909.09436}
  (\bibinfo{year}{2019}).
\newblock
\urldef\tempurl%
\url{http://arxiv.org/abs/1909.09436}
\showURL{%
\tempurl}


\bibitem[\protect\citeauthoryear{Kudo and Richardson}{Kudo and
  Richardson}{2018}]%
        {Kudo:sentencePiece}
\bibfield{author}{\bibinfo{person}{Taku Kudo} {and} \bibinfo{person}{John
  Richardson}.} \bibinfo{year}{2018}\natexlab{}.
\newblock \showarticletitle{SentencePiece: {A} simple and language independent
  subword tokenizer and detokenizer for Neural Text Processing}.
\newblock \bibinfo{journal}{\emph{CoRR}} (\bibinfo{year}{2018}).
\newblock
\showeprint[arxiv]{1808.06226}


\bibitem[\protect\citeauthoryear{Lewis, Lin, Sadowski, Zhu, Ou, and
  Whitehead}{Lewis et~al\mbox{.}}{2013}]%
        {lewis2013does}
\bibfield{author}{\bibinfo{person}{Chris Lewis}, \bibinfo{person}{Zhongpeng
  Lin}, \bibinfo{person}{Caitlin Sadowski}, \bibinfo{person}{Xiaoyan Zhu},
  \bibinfo{person}{Rong Ou}, {and} \bibinfo{person}{E~James Whitehead}.}
  \bibinfo{year}{2013}\natexlab{}.
\newblock \showarticletitle{Does bug prediction support human developers?
  findings from a google case study}. In \bibinfo{booktitle}{\emph{2013 35th
  International Conference on Software Engineering (ICSE)}}. IEEE,
  \bibinfo{pages}{372--381}.
\newblock


\bibitem[\protect\citeauthoryear{Mastropaolo, Scalabrino, Cooper, Palacio,
  Poshyvanyk, Oliveto, and Bavota}{Mastropaolo et~al\mbox{.}}{2021}]%
        {mastropaolo2021studying}
\bibfield{author}{\bibinfo{person}{Antonio Mastropaolo},
  \bibinfo{person}{Simone Scalabrino}, \bibinfo{person}{Nathan Cooper},
  \bibinfo{person}{David~Nader Palacio}, \bibinfo{person}{Denys Poshyvanyk},
  \bibinfo{person}{Rocco Oliveto}, {and} \bibinfo{person}{Gabriele Bavota}.}
  \bibinfo{year}{2021}\natexlab{}.
\newblock \showarticletitle{Studying the Usage of Text-To-Text Transfer
  Transformer to Support Code-Related Tasks}. In \bibinfo{booktitle}{\emph{2021
  IEEE/ACM 43rd International Conference on Software Engineering (ICSE)}}.
  IEEE, \bibinfo{pages}{336--347}.
\newblock


\bibitem[\protect\citeauthoryear{McIntosh, Kamei, Adams, and Hassan}{McIntosh
  et~al\mbox{.}}{2014}]%
        {McIntosh:msr2014}
\bibfield{author}{\bibinfo{person}{Shane McIntosh}, \bibinfo{person}{Yasutaka
  Kamei}, \bibinfo{person}{Bram Adams}, {and} \bibinfo{person}{Ahmed~E.
  Hassan}.} \bibinfo{year}{2014}\natexlab{}.
\newblock \showarticletitle{The Impact of Code Review Coverage and Code Review
  Participation on Software Quality: A Case Study of the Qt, VTK, and ITK
  Projects}. In \bibinfo{booktitle}{\emph{Proceedings of the 11th Working
  Conference on Mining Software Repositories}} \emph{(\bibinfo{series}{MSR
  2014})}. \bibinfo{pages}{192--201}.
\newblock


\bibitem[\protect\citeauthoryear{McNemar}{McNemar}{1947}]%
        {mcnemar}
\bibfield{author}{\bibinfo{person}{Quinn McNemar}.}
  \bibinfo{year}{1947}\natexlab{}.
\newblock \showarticletitle{Note on the sampling error of the difference
  between correlated proportions or percentages}.
\newblock \bibinfo{journal}{\emph{Psychometrika}} \bibinfo{volume}{12},
  \bibinfo{number}{2} (\bibinfo{year}{1947}), \bibinfo{pages}{153--157}.
\newblock


\bibitem[\protect\citeauthoryear{Morales, McIntosh, and Khomh}{Morales
  et~al\mbox{.}}{2015}]%
        {morales2015saner}
\bibfield{author}{\bibinfo{person}{Rodrigo Morales}, \bibinfo{person}{Shane
  McIntosh}, {and} \bibinfo{person}{Foutse Khomh}.}
  \bibinfo{year}{2015}\natexlab{}.
\newblock \showarticletitle{Do Code Review Practices Impact Design Quality? A
  Case Study of the Qt, VTK, and ITK Projects}. In
  \bibinfo{booktitle}{\emph{Proc. of the 22nd Int'l Conf. on Software Analysis,
  Evolution, and Reengineering (SANER)}}. \bibinfo{pages}{171--180}.
\newblock


\bibitem[\protect\citeauthoryear{Papineni, Roukos, Ward, and Zhu}{Papineni
  et~al\mbox{.}}{2002}]%
        {Papineni:2002}
\bibfield{author}{\bibinfo{person}{Kishore Papineni}, \bibinfo{person}{Salim
  Roukos}, \bibinfo{person}{Todd Ward}, {and} \bibinfo{person}{Wei-Jing Zhu}.}
  \bibinfo{year}{2002}\natexlab{}.
\newblock \showarticletitle{BLEU: A Method for Automatic Evaluation of Machine
  Translation}. In \bibinfo{booktitle}{\emph{Proceedings of the 40th Annual
  Meeting on Association for Computational Linguistics}}
  \emph{(\bibinfo{series}{ACL '02})}. \bibinfo{pages}{311--318}.
\newblock


\bibitem[\protect\citeauthoryear{Pascarella, Palomba, and Bacchelli}{Pascarella
  et~al\mbox{.}}{2019}]%
        {pascarella2019fine}
\bibfield{author}{\bibinfo{person}{Luca Pascarella}, \bibinfo{person}{Fabio
  Palomba}, {and} \bibinfo{person}{Alberto Bacchelli}.}
  \bibinfo{year}{2019}\natexlab{}.
\newblock \showarticletitle{Fine-grained just-in-time defect prediction}.
\newblock \bibinfo{journal}{\emph{Journal of Systems and Software}}
  \bibinfo{volume}{150} (\bibinfo{year}{2019}), \bibinfo{pages}{22--36}.
\newblock


\bibitem[\protect\citeauthoryear{Pascarella, Spadini, Palomba, Bruntink, and
  Bacchelli}{Pascarella et~al\mbox{.}}{2018}]%
        {pascarella2018information}
\bibfield{author}{\bibinfo{person}{Luca Pascarella}, \bibinfo{person}{Davide
  Spadini}, \bibinfo{person}{Fabio Palomba}, \bibinfo{person}{Magiel Bruntink},
  {and} \bibinfo{person}{Alberto Bacchelli}.} \bibinfo{year}{2018}\natexlab{}.
\newblock \showarticletitle{Information needs in contemporary code review}.
\newblock \bibinfo{journal}{\emph{Proceedings of the ACM on Human-Computer
  Interaction}} \bibinfo{volume}{2}, \bibinfo{number}{CSCW}
  (\bibinfo{year}{2018}), \bibinfo{pages}{1--27}.
\newblock


\bibitem[\protect\citeauthoryear{Raffel, Shazeer, Roberts, Lee, Narang, Matena,
  Zhou, Li, and Liu}{Raffel et~al\mbox{.}}{2020}]%
        {raffel2019exploring}
\bibfield{author}{\bibinfo{person}{Colin Raffel}, \bibinfo{person}{Noam
  Shazeer}, \bibinfo{person}{Adam Roberts}, \bibinfo{person}{Katherine Lee},
  \bibinfo{person}{Sharan Narang}, \bibinfo{person}{Michael Matena},
  \bibinfo{person}{Yanqi Zhou}, \bibinfo{person}{Wei Li}, {and}
  \bibinfo{person}{Peter~J. Liu}.} \bibinfo{year}{2020}\natexlab{}.
\newblock \showarticletitle{Exploring the Limits of Transfer Learning with a
  Unified Text-to-Text Transformer}.
\newblock \bibinfo{journal}{\emph{Journal of Machine Learning Research}}
  \bibinfo{volume}{21}, \bibinfo{number}{140} (\bibinfo{year}{2020}),
  \bibinfo{pages}{1--67}.
\newblock
\urldef\tempurl%
\url{http://jmlr.org/papers/v21/20-074.html}
\showURL{%
\tempurl}


\bibitem[\protect\citeauthoryear{Raychev, Vechev, and Yahav}{Raychev
  et~al\mbox{.}}{2014}]%
        {Raychev:pldi2014}
\bibfield{author}{\bibinfo{person}{Veselin Raychev}, \bibinfo{person}{Martin
  Vechev}, {and} \bibinfo{person}{Eran Yahav}.}
  \bibinfo{year}{2014}\natexlab{}.
\newblock \showarticletitle{Code Completion with Statistical Language Models}.
  In \bibinfo{booktitle}{\emph{Proceedings of the 35th ACM SIGPLAN Conference
  on Programming Language Design and Implementation}}
  \emph{(\bibinfo{series}{PLDI '14})}. \bibinfo{publisher}{ACM},
  \bibinfo{pages}{419--428}.
\newblock


\bibitem[\protect\citeauthoryear{Ren, Guo, Lu, Zhou, Liu, Tang, Sundaresan,
  Zhou, Blanco, and Ma}{Ren et~al\mbox{.}}{2020}]%
        {ren2020codebleu}
\bibfield{author}{\bibinfo{person}{Shuo Ren}, \bibinfo{person}{Daya Guo},
  \bibinfo{person}{Shuai Lu}, \bibinfo{person}{Long Zhou},
  \bibinfo{person}{Shujie Liu}, \bibinfo{person}{Duyu Tang},
  \bibinfo{person}{Neel Sundaresan}, \bibinfo{person}{Ming Zhou},
  \bibinfo{person}{Ambrosio Blanco}, {and} \bibinfo{person}{Shuai Ma}.}
  \bibinfo{year}{2020}\natexlab{}.
\newblock \bibinfo{title}{CodeBLEU: a Method for Automatic Evaluation of Code
  Synthesis}.
\newblock
\newblock
\showeprint[arxiv]{2009.10297}~[cs.SE]


\bibitem[\protect\citeauthoryear{Rigby and Bird}{Rigby and Bird}{2013}]%
        {Rigby:fse2013}
\bibfield{author}{\bibinfo{person}{Peter~C. Rigby} {and}
  \bibinfo{person}{Christian Bird}.} \bibinfo{year}{2013}\natexlab{}.
\newblock \showarticletitle{Convergent Contemporary Software Peer Review
  Practices}. In \bibinfo{booktitle}{\emph{Proceedings of the 2013 9th Joint
  Meeting on Foundations of Software Engineering}}
  \emph{(\bibinfo{series}{ESEC/FSE 2013})}. \bibinfo{pages}{202--212}.
\newblock


\bibitem[\protect\citeauthoryear{Rigby, German, Cowen, and Storey}{Rigby
  et~al\mbox{.}}{2014}]%
        {Rigby:tosem2014}
\bibfield{author}{\bibinfo{person}{Peter~C. Rigby}, \bibinfo{person}{Daniel~M.
  German}, \bibinfo{person}{Laura Cowen}, {and} \bibinfo{person}{Margaret-Anne
  Storey}.} \bibinfo{year}{2014}\natexlab{}.
\newblock \showarticletitle{Peer Review on Open-Source Software Projects:
  Parameters, Statistical Models, and Theory}.
\newblock \bibinfo{journal}{\emph{ACM Trans. Softw. Eng. Methodol.}}
  \bibinfo{volume}{23}, \bibinfo{number}{4} (\bibinfo{year}{2014}).
\newblock


\bibitem[\protect\citeauthoryear{Sadowski, S\"{o}derberg, Church, Sipko, and
  Bacchelli}{Sadowski et~al\mbox{.}}{2018}]%
        {Caitlin:icse2018}
\bibfield{author}{\bibinfo{person}{Caitlin Sadowski}, \bibinfo{person}{Emma
  S\"{o}derberg}, \bibinfo{person}{Luke Church}, \bibinfo{person}{Michal
  Sipko}, {and} \bibinfo{person}{Alberto Bacchelli}.}
  \bibinfo{year}{2018}\natexlab{}.
\newblock \showarticletitle{Modern Code Review: A Case Study at Google}. In
  \bibinfo{booktitle}{\emph{Proceedings of the 40th International Conference on
  Software Engineering: Software Engineering in Practice}}
  \emph{(\bibinfo{series}{ICSE-SEIP '18})}. \bibinfo{pages}{181?190}.
\newblock


\bibitem[\protect\citeauthoryear{Shi, Li, Lo, Thung, and Huo}{Shi
  et~al\mbox{.}}{2019}]%
        {shi2019automatic}
\bibfield{author}{\bibinfo{person}{Shu-Ting Shi}, \bibinfo{person}{Ming Li},
  \bibinfo{person}{David Lo}, \bibinfo{person}{Ferdian Thung}, {and}
  \bibinfo{person}{Xuan Huo}.} \bibinfo{year}{2019}\natexlab{}.
\newblock \showarticletitle{Automatic code review by learning the revision of
  source code}. In \bibinfo{booktitle}{\emph{Proceedings of the AAAI Conference
  on Artificial Intelligence}}, Vol.~\bibinfo{volume}{33}.
  \bibinfo{pages}{4910--4917}.
\newblock


\bibitem[\protect\citeauthoryear{Singh, Sekar, Stolee, and Johnson}{Singh
  et~al\mbox{.}}{2017}]%
        {singh2017evaluating}
\bibfield{author}{\bibinfo{person}{Devarshi Singh},
  \bibinfo{person}{Varun~Ramachandra Sekar}, \bibinfo{person}{Kathryn~T
  Stolee}, {and} \bibinfo{person}{Brittany Johnson}.}
  \bibinfo{year}{2017}\natexlab{}.
\newblock \showarticletitle{Evaluating how static analysis tools can reduce
  code review effort}. In \bibinfo{booktitle}{\emph{2017 IEEE Symposium on
  Visual Languages and Human-Centric Computing (VL/HCC)}}. IEEE,
  \bibinfo{pages}{101--105}.
\newblock


\bibitem[\protect\citeauthoryear{Spadini, Palomba, Baum, Hanenberg, Bruntink,
  and Bacchelli}{Spadini et~al\mbox{.}}{2019}]%
        {spadini2019test}
\bibfield{author}{\bibinfo{person}{Davide Spadini}, \bibinfo{person}{Fabio
  Palomba}, \bibinfo{person}{Tobias Baum}, \bibinfo{person}{Stefan Hanenberg},
  \bibinfo{person}{Magiel Bruntink}, {and} \bibinfo{person}{Alberto
  Bacchelli}.} \bibinfo{year}{2019}\natexlab{}.
\newblock \showarticletitle{Test-driven code review: an empirical study}. In
  \bibinfo{booktitle}{\emph{2019 IEEE/ACM 41st International Conference on
  Software Engineering (ICSE)}}. IEEE, \bibinfo{pages}{1061--1072}.
\newblock


\bibitem[\protect\citeauthoryear{Tufano, Drain, Svyatkovskiy, Deng, and
  Sundaresan}{Tufano et~al\mbox{.}}{2020}]%
        {tufano2020unitTest}
\bibfield{author}{\bibinfo{person}{Michele Tufano}, \bibinfo{person}{Dawn
  Drain}, \bibinfo{person}{Alexey Svyatkovskiy}, \bibinfo{person}{Shao~Kun
  Deng}, {and} \bibinfo{person}{Neel Sundaresan}.}
  \bibinfo{year}{2020}\natexlab{}.
\newblock \showarticletitle{Unit Test Case Generation with Transformers}.
\newblock \bibinfo{journal}{\emph{CoRR}}  \bibinfo{volume}{abs/2009.05617}
  (\bibinfo{year}{2020}).
\newblock
\urldef\tempurl%
\url{https://arxiv.org/abs/2009.05617}
\showURL{%
\tempurl}


\bibitem[\protect\citeauthoryear{Tufano, Watson, Bavota, Penta, White, and
  Poshyvanyk}{Tufano et~al\mbox{.}}{2019}]%
        {Tufano:tosem2019}
\bibfield{author}{\bibinfo{person}{Michele Tufano}, \bibinfo{person}{Cody
  Watson}, \bibinfo{person}{Gabriele Bavota}, \bibinfo{person}{Massimiliano~Di
  Penta}, \bibinfo{person}{Martin White}, {and} \bibinfo{person}{Denys
  Poshyvanyk}.} \bibinfo{year}{2019}\natexlab{}.
\newblock \showarticletitle{An Empirical Study on Learning Bug-Fixing Patches
  in the Wild via Neural Machine Translation}.
\newblock \bibinfo{journal}{\emph{{ACM} Trans. Softw. Eng. Methodol.}}
  \bibinfo{volume}{28}, \bibinfo{number}{4} (\bibinfo{year}{2019}),
  \bibinfo{pages}{19:1--19:29}.
\newblock


\bibitem[\protect\citeauthoryear{Tufano, Pascarella, Tufano, Poshyvanyk, and
  Bavota}{Tufano et~al\mbox{.}}{2021}]%
        {Tufano:icse2021}
\bibfield{author}{\bibinfo{person}{Rosalia Tufano}, \bibinfo{person}{Luca
  Pascarella}, \bibinfo{person}{Michele Tufano}, \bibinfo{person}{Denys
  Poshyvanyk}, {and} \bibinfo{person}{Gabriele Bavota}.}
  \bibinfo{year}{2021}\natexlab{}.
\newblock \showarticletitle{Towards Automating Code Review Activities}. In
  \bibinfo{booktitle}{\emph{43rd International Conference on Software
  Engineering, {ICSE}'21}}.
\newblock
\urldef\tempurl%
\url{https://arxiv.org/abs/2101.02518}
\showURL{%
\tempurl}


\bibitem[\protect\citeauthoryear{Tymchuk, Mocci, and Lanza}{Tymchuk
  et~al\mbox{.}}{2015}]%
        {tymchuk2015code}
\bibfield{author}{\bibinfo{person}{Yuriy Tymchuk}, \bibinfo{person}{Andrea
  Mocci}, {and} \bibinfo{person}{Michele Lanza}.}
  \bibinfo{year}{2015}\natexlab{}.
\newblock \showarticletitle{Code review: Veni, vidi, vici}. In
  \bibinfo{booktitle}{\emph{2015 IEEE 22nd International Conference on Software
  Analysis, Evolution, and Reengineering (SANER)}}. IEEE,
  \bibinfo{pages}{151--160}.
\newblock


\bibitem[\protect\citeauthoryear{Vaswani, Shazeer, Parmar, Uszkoreit, Jones,
  Gomez, Kaiser, and Polosukhin}{Vaswani et~al\mbox{.}}{2017}]%
        {vaswani2017attention}
\bibfield{author}{\bibinfo{person}{Ashish Vaswani}, \bibinfo{person}{Noam
  Shazeer}, \bibinfo{person}{Niki Parmar}, \bibinfo{person}{Jakob Uszkoreit},
  \bibinfo{person}{Llion Jones}, \bibinfo{person}{Aidan~N Gomez},
  \bibinfo{person}{{\L}ukasz Kaiser}, {and} \bibinfo{person}{Illia
  Polosukhin}.} \bibinfo{year}{2017}\natexlab{}.
\newblock \showarticletitle{Attention is all you need}. In
  \bibinfo{booktitle}{\emph{Advances in neural information processing
  systems}}. \bibinfo{pages}{5998--6008}.
\newblock


\bibitem[\protect\citeauthoryear{Watson, Cooper, Palacio, Moran, and
  Poshyvanyk}{Watson et~al\mbox{.}}{2020a}]%
        {watson2020systematic}
\bibfield{author}{\bibinfo{person}{Cody Watson}, \bibinfo{person}{Nathan
  Cooper}, \bibinfo{person}{David~Nader Palacio}, \bibinfo{person}{Kevin
  Moran}, {and} \bibinfo{person}{Denys Poshyvanyk}.}
  \bibinfo{year}{2020}\natexlab{a}.
\newblock \showarticletitle{A Systematic Literature Review on the Use of Deep
  Learning in Software Engineering Research}.
\newblock \bibinfo{journal}{\emph{arXiv preprint arXiv:2009.06520}}
  (\bibinfo{year}{2020}).
\newblock


\bibitem[\protect\citeauthoryear{Watson, Tufano, Moran, Bavota, and
  Poshyvanyk}{Watson et~al\mbox{.}}{2020b}]%
        {Watson:icse2020}
\bibfield{author}{\bibinfo{person}{Cody Watson}, \bibinfo{person}{Michele
  Tufano}, \bibinfo{person}{Kevin Moran}, \bibinfo{person}{Gabriele Bavota},
  {and} \bibinfo{person}{Denys Poshyvanyk}.} \bibinfo{year}{2020}\natexlab{b}.
\newblock \showarticletitle{On learning meaningful assert statements for unit
  test cases}. In \bibinfo{booktitle}{\emph{{ICSE} '20: 42nd International
  Conference on Software Engineering, Seoul, South Korea, 27 June - 19 July,
  2020}}, \bibfield{editor}{\bibinfo{person}{Gregg Rothermel} {and}
  \bibinfo{person}{Doo{-}Hwan Bae}} (Eds.). \bibinfo{publisher}{{ACM}},
  \bibinfo{pages}{1398--1409}.
\newblock


\bibitem[\protect\citeauthoryear{Wattanakriengkrai, Thongtanunam,
  Tantithamthavorn, Hata, and Matsumoto}{Wattanakriengkrai
  et~al\mbox{.}}{2020}]%
        {Wattanakriengkrai2020linelevel}
\bibfield{author}{\bibinfo{person}{Supatsara Wattanakriengkrai},
  \bibinfo{person}{Patanamon Thongtanunam}, \bibinfo{person}{Chakkrit
  Tantithamthavorn}, \bibinfo{person}{Hideaki Hata}, {and}
  \bibinfo{person}{Kenichi Matsumoto}.} \bibinfo{year}{2020}\natexlab{}.
\newblock \showarticletitle{Predicting Defective Lines Using a Model-Agnostic
  Technique}.
\newblock \bibinfo{journal}{\emph{CoRR}} (\bibinfo{year}{2020}).
\newblock


\end{thebibliography}

\end{document}